\documentclass[superscriptaddress,pre,onecolumn]{revtex4-2}
\usepackage{amsmath}
\usepackage{amssymb}
\usepackage{amscd}
\usepackage{graphicx}
\usepackage{bbm}
\usepackage{dsfont}
\usepackage{datetime}
\longdate

\begin{document}
\title{Active Nematic Multipoles: Flow Responses and the Dynamics of Defects and Colloids}
\author{Alexander J.H. Houston}
\affiliation{Department of Physics, Gibbet Hill Road, University of Warwick, Coventry, CV4 7AL, United Kingdom.}
\author{Gareth P. Alexander}
\email{G.P.Alexander@warwick.ac.uk}
\affiliation{Department of Physics, Gibbet Hill Road, University of Warwick, Coventry, CV4 7AL, United Kingdom.}
\affiliation{Centre for Complexity Science, Zeeman Building, University of Warwick, Coventry, CV4 7AL, United Kingdom.}

\date{\today}

\begin{abstract}
We introduce a general description of localised distortions in active nematics using the framework of active nematic multipoles. We give the Stokesian flows for arbitrary multipoles in terms of differentiation of a fundamental flow response and describe them explicitly up to quadrupole order. We also present the response in terms of the net active force and torque associated to the multipole. This allows the identification of the dipolar and quadrupolar distortions that generate self-propulsion and self-rotation respectively and serves as a guide for the design of arbitrary flow responses. Our results can be applied to both defect loops in three-dimensional active nematics and to systems with colloidal inclusions. They reveal the geometry-dependence of the self-dynamics of defect loops and provide insights into how colloids might be designed to achieve propulsive or rotational dynamics, and more generally for the extraction of work from active nematics. Finally, we extend our analysis also to two dimensions and to systems with chiral active stresses.
\end{abstract}
\maketitle

\section{Introduction}
\label{sec:intro}

Active liquid crystals model a wide range of materials, both biological and synthetic ~\cite{ramaswamy2010mechanics,marchetti2013hydrodynamics,doostmohammadi2018active}, including cell monolayers~\cite{duclos2017topological}, tissues~\cite{saw2017topological}, bacteria in liquid crystalline environments~\cite{zhou2014living} and bacterial suspensions~\cite{wensink2012meso}, and synthetic suspensions of microtubules~\cite{sanchez2012spontaneous}. Nematic and polar phases have been the focus of attention but smectic~\cite{adhyapak2013live,chen_toner2013}, cholesteric~\cite{whitfield2017hydrodynamic,kole2021layered} and hexatic~\cite{maitra2020} phases have also been considered. 
Key features and motifs of the active nematic state include self-propelled topological defects~\cite{narayan2007,giomi2013defect,giomi2014defect}, spontaneous flows and vortices, and on how these may be controlled through boundary conditions, confinement \cite{shendruk2017dancing,norton2018insensitivity,opathalage2019self}, external fields, geometry or topology. Active defects, in particular, have been related to processes of apoptosis in epithelial sheets~\cite{saw2017topological}, tissue dynamics, bacterial spreading and biofilm formation, and morphogenesis in {\it Hydra}~\cite{maroudas-sacks2021}. 

In three-dimensional active nematics the fundamental excitations are defect loops and system-spanning lines~\cite{duclos2020topological,vcopar2019topology}. The defect loops actively self-propel~\cite{binysh2020three}, and self-orient~\cite{houston2022defect}, in addition to undergoing deformations in shape. Their finite extent means that they represent localised distortions to the nematic director, on scales larger than their size, and this facilitates a description through elastic multipoles~\cite{houston2022defect}. It also invites comparison with colloidal inclusions in passive liquid crystals, which create localised realignments of the director and act as elastic multipoles~\cite{poulin1997novel,stark2001physics,muvsevivc2017liquid}. These multipole distortions mediate interactions between colloids and allow for a means of controlling both the colloidal inclusions and the host material. For instance, they facilitate self-assembly and the formation of metamaterials~\cite{muvsevivc2006two,ravnik2013confined}, and enable novel control of topological defects~\cite{ravnik2007,tkalec2011reconfigurable,muvsevivc2017liquid}. While there have been studies of active nematic droplets in a host passive liquid crystal~\cite{guillamat2018,hardouin2019}, colloidal inclusions in host active nematics have not been looked at previously.  

The multipole approach to describing colloidal inclusions and localised director distortions in general, offers an equally fruitful paradigm in active nematics. Here, we present a generic analysis of the active flows generated by multipole director distortions in an active nematic and predict that the presence of colloids transforms their behaviour similarly to the passive case. These active multipole flows represent the responses of the active nematic both to localised features, such as defect loops, and to colloidal inclusions. This allows us to identify those distortions which produce directed or rotational flows and show that such distortions may be naturally induced by colloids. We also characterise the response in terms of the active forces and torques that they induce. This general connection can serve as a guide for using colloidal inclusions as a means to control active nematics, or how to design them to engineer a desired response, or extract work. The properties of inclusions have been studied in scalar active matter \cite{baek2018generic}, as have active droplets in passive nematics \cite{rajabi2020directional}, but while there have been specific demonstrations of propulsive colloids \cite{loewe2022passive,yao2022topological} the general responses of inclusions in active nematics have not previously been considered. Understanding how such responses relate to local manipulations and molecular fields in active nematics will bring both fundamental insights and the potential for control of active metamaterials.

The remainder of this paper is structured as follows. In Section 2 we briefly review the equations of active nematohydrodynamics and describe the regime in which our linear multipole approach applies. In Section 3 we present these multipoles as complex derivatives acting on $1/r$, showing how this naturally elucidates their symmetries. In Section 4 we show that the linear active response to a harmonic distortion is generated by the same complex derivatives acting on fundamental flow and pressure solutions and highlight certain examples that illustrate the self-propulsive and rotational dynamics that can arise. We then show in Section 5 that these phenomenological responses can be discerned from integrals of the active stress, allowing the identification of the distortion which produces propulsion along or rotation about a given axis. Sections 6 and 7 contain extensions of our approach, first to two-dimensional systems and then to those with chiral active stresses. Section 8 gives a discussion and summary.

\section{Hydrodynamics of active nematics}
\label{sec:hydro}
We summarise the hydrodynamics of active nematics as described by their director field ${\bf n}$ and fluid velocity ${\bf u}$. The fluid flow satisfies the continuity $\partial_i u_i = 0$ and Stokes $\partial_j \sigma_{ij} = 0$ equations, with stress tensor~\cite{ramaswamy2010mechanics,marchetti2013hydrodynamics,doostmohammadi2018active}  
\begin{equation}
    \begin{split}
        \sigma_{ij} &= - p \delta_{ij} + 2\mu D_{ij} + \frac{\nu}{2} \bigl( n_i h_j + h_i n_j \bigr) + \frac{1}{2} \bigl( n_i h_j - h_i n_j \bigr) + \sigma^{\textrm{E}}_{ij} - \zeta n_i n_j .
    \end{split}
\end{equation}
Here, $p$ is the pressure, $\mu$ is the viscosity, $D_{ij} = \frac{1}{2} (\partial_i u_j + \partial_j u_i)$ is the symmetric part of the velocity gradients, $\nu$ is the flow alignment parameter, $h_i = - \delta F/\delta n_i$ is the molecular field associated with the Frank free energy $F$, $\sigma^{\textrm{E}}_{ij}$ is the Ericksen stress, and $\zeta$ is the magnitude of the activity. The active nematic is extensile when $\zeta > 0$ and contractile when $\zeta < 0$. The director field satisfies the relaxational equation 
\begin{equation}
    \begin{split}
        \partial_t n_i + u_j \partial_j n_i + \Omega_{ij} n_j & = \frac{1}{\gamma} h_i- \nu \bigl[ D_{ij} n_j - n_i (n_j D_{jk} n_k) \bigr] ,
    \end{split}
\end{equation}
where $\gamma$ is a rotational viscosity and $\Omega_{ij} = \frac{1}{2} (\partial_i u_j - \partial_j u_i)$ is the antisymmetric part of the velocity gradients. We adopt a one-elastic-constant approximation for the Frank free energy \cite{frank1958liquid}
\begin{equation}
F = \int \frac{K}{2} \bigl( \partial_i n_j \bigr) \bigl( \partial_i n_j \bigr) \,dV ,
\end{equation}
for which the molecular field is $h_i = K \bigl( \nabla^2 n_i - n_i n_j \nabla^2 n_j \bigr)$ and the Ericksen stress is $\sigma^{\textrm{E}}_{ij} = - K \partial_i n_k \,\partial_j n_k$.

An often-used analytical approximation is to consider the active flows generated by an equilibrium director field. This approximation has been used previously in the theoretical description of the active flows generated by defects in both two~\cite{giomi2014defect,angheluta2021role} and three dimensions~\cite{binysh2020three}, including on curved surfaces~\cite{khoromskaia2017vortex}, and in active turbulence~\cite{alert2020universal}. It may be thought of in terms of a limit of weak activity, however, even when the activity is strong enough to generate defects, their structure may still be close to that of equilibrium defects and the approximation remain good and the comparison of active defect motion and flows described in this way with full numerical simulations suggests that this is at least qualitatively the case. The equations can then be reduced to ${\bf h} = {\bf 0}$ for the director field and the Stokes equation
\begin{equation}
- \nabla p + \mu \nabla^2 {\bf u} = \zeta \nabla \cdot \bigl( {\bf nn} \bigr) ,
\label{eq:active_Stokes}
\end{equation}
for the active flow. Here we have neglected the Ericksen stress since for an equilibrium director field it can be balanced by a contribution to the pressure (representing nematic hydrostatic equilibrium). 

\begin{figure}[t]
    \centering
    \includegraphics[width=0.35\textwidth]{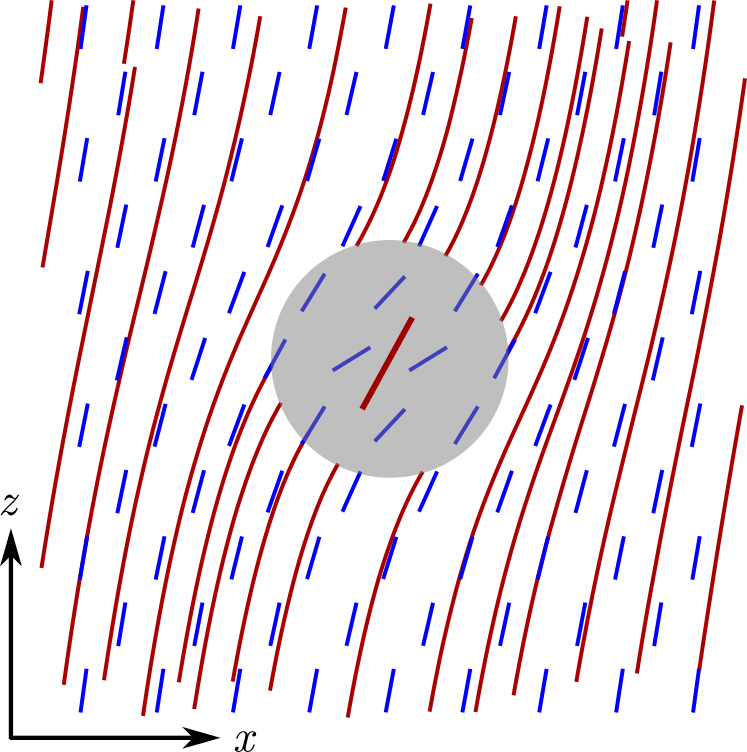}
    \caption{Comparison of the exact director field (red streamlines) and linearised multipole approximation (blue rods) for the most slowly decaying monopole distortion. This is produced by uniformly rotating the director by an angle $\theta_0$ within a spherical volume of radius $a$, indicated by the grey disc; the alignment inside the sphere is indicated by the thick red line. The figure shows only the $xz$-plane in which the director rotates and in which the comparison is most strict.}
    \label{fig:monopole}
\end{figure}

We limit our analysis to director fields that can be linearised around a (locally) uniformly aligned state, ${\bf n} = {\bf e}_z + \delta {\bf n}$, with $\delta {\bf n} \cdot {\bf e}_z = 0$, for which the equations reduce to 
\begin{gather}
\nabla^2 \delta {\bf n} = 0 , \label{eq:director_Laplace} \\
\nabla \cdot {\bf u} = 0 , \label{eq:continuity} \\
- \nabla p + \mu \nabla^2 {\bf u} = \zeta \bigl[ {\bf e}_z \bigl( \nabla \cdot \delta {\bf n} \bigr) + \partial_z \delta {\bf n} \bigr] . \label{eq:active_Stokes_linearised}
\end{gather}
These correspond to elastic multipole states in the director field, which are often thought of as an asymptotic description, however, they provide a close approximation even at only moderate distances outside a `core' region that is the source of the multipole. To illustrate this we show in Fig.~\ref{fig:monopole} a comparison between the exact director field (red streamlines) and linear multipole approximation (blue rods) for the most slowly varying monopole distortion created by uniformly rotating the director by an angle $\theta_0$ within a sphere of radius $a$. The agreement is close anywhere outside the sphere and only deviates significantly in the near-field region inside it. This is relevant to the active system as it is well-known that the uniformly aligned active nematic state is fundamentally unstable~\cite{simha2002hydrodynamic} and active nematics are turbulent on large enough scales. Our solutions should be interpreted as describing the behaviour on intermediate scales, larger than the core structure of the source but smaller than the scale on which turbulence takes over.

\section{Multipole director distortions}
\label{Chap3:sec:Multipole director distortions}

\begin{figure}
\centering
\includegraphics[width=0.92\textwidth]{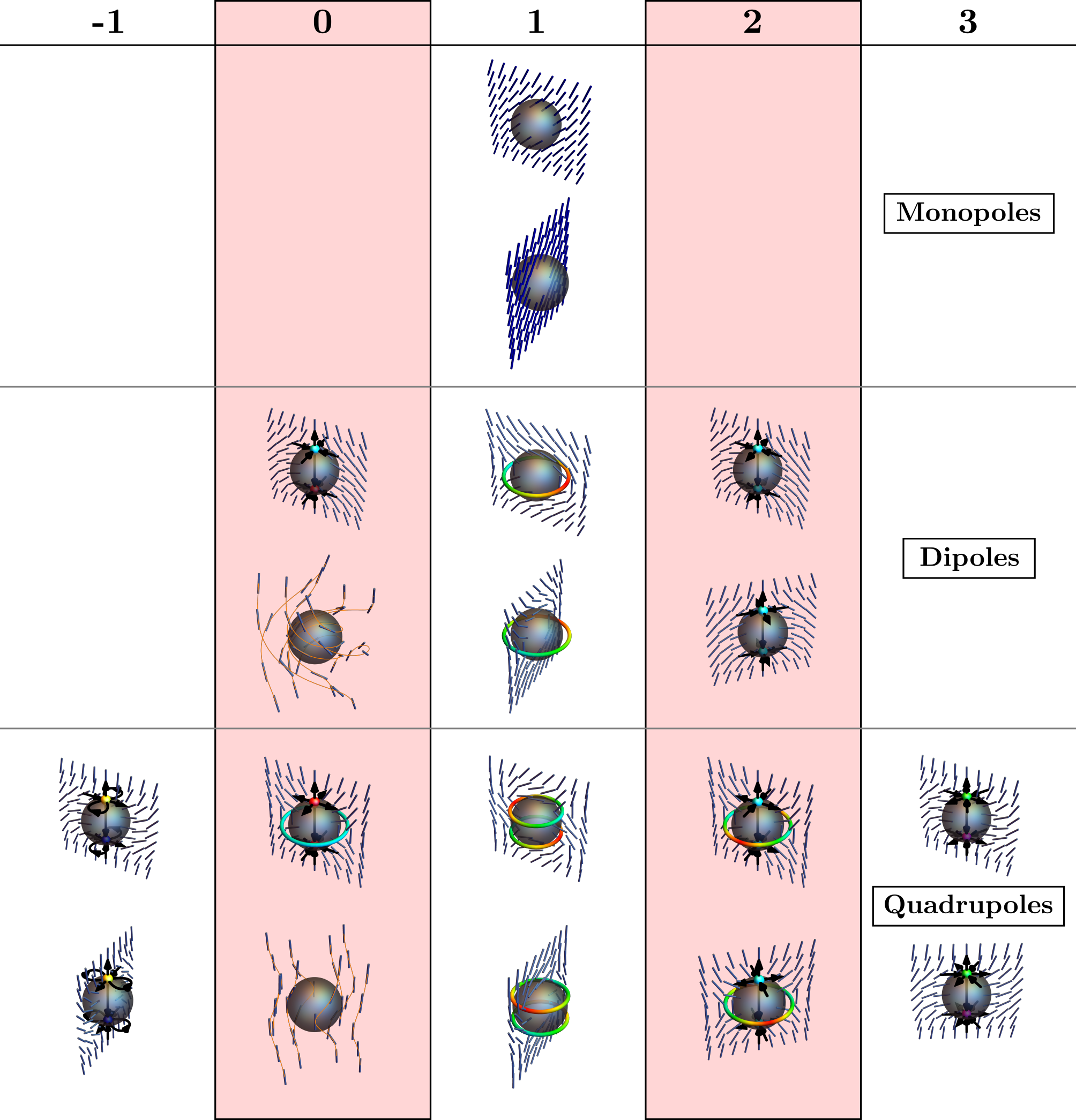}
    \caption[The multipolar director distortions up to quadrupole order.]{The multipolar director distortions up to quadrupole order. The director is shown on a planar cross-section as blue rods, along with a topological skeleton corresponding to the spherical harmonic, where appropriate. Defect loops are coloured according to wedge (blue) or twist (red-green) type and the charge of point defects is indicated through the use of opposing colour pairs: red ($+1$) and cyan ($-1$), yellow ($+2$) and blue ($-2$), and green ($+3$) and magenta ($-3$). Their charge is further indicated by a local decoration of the director with an orientation, indicated by black arrows. Each multipole order is classified into vertical pairs according to the spin of the distortion. For the chiral multipoles, the visualisation instead shows the director along some of its integral curves (orange).}
    \label{fig:3DNematicMultipoles}
\end{figure}

In this section, we describe the multipole director fields satisfying~\eqref{eq:director_Laplace}. The far-field orientation ${\bf e}_z$ gives a splitting of directions in space into those parallel and perpendicular to it. We complexify the perpendicular plane to give the decomposition as $\mathbb{R}^3 \cong \mathbb{C}\oplus\mathbb{R}$ and convert the director deformation $\delta {\bf n}$ to the complex form $\delta n = \delta n_x + i \delta n_y$. The real and imaginary parts of $\delta n$ are harmonic, meaning that at order $l$ they may be expressed as spherical harmonics $1/r^{l+1}Y_m^l$ or, as we shall do, as $l$ derivatives of $1/r$~\cite{maxwell1873treatise,dennis2004canonical,arnold1996topological}. These order $l$ multipole solutions form a $2(2l+1)$-real-dimensional vector space. Associated to the $\mathbb{C}\oplus\mathbb{R}$ splitting is a local symmetry group isomorphic to $U(1)$, preserving ${\bf e}_z$, whose irreducible representations provide a natural basis for the vector space of multipoles at each order. We write the complex derivatives on $\mathbb{C}$ as $\partial_w=\frac{1}{2}(\partial_x-i\partial_y)$ and $\partial_{\bar{w}}=\frac{1}{2}(\partial_x+i\partial_y)$ in terms of which the director deformation can be written 
\begin{equation}
    \delta n = \sum_{l=0}^{\infty} \sum_{m=-l}^l q_{lm} \,a^{l+1} \,\partial^m_{\bar{w}} \partial^{l-m}_z \,\frac{1}{r} ,
\label{ComplexHarmonics}
\end{equation}
where $q_{lm}$ are complex coefficients and $a$ is a characteristic length scale of the multipole, as might be set by the radius of a colloid. For compactness of notation it is to be understood that when $m$ is negative $\partial^m_{\bar{w}}$ represents $\partial^{|m|}_w$. The index $m$ denotes the topological charge of the phase winding of the spherical harmonic. This gives the spin of the corresponding vector field as $1-m$, where the $1$ is due to a vector ($\delta {\bf n}$ or $\delta n$) being a spin-$1$ object. The multipoles at order $l$ therefore have spins that range from $1-l$ to $1+l$. They are illustrated up to quadrupole order in Fig.~\ref{fig:3DNematicMultipoles}, along with a representation in terms of topological defects which we shall elaborate upon shortly. The structure of Fig.~\ref{fig:3DNematicMultipoles} is such that differentiation maps the distortions of one order to the next, with $\partial_z$ leaving the distortion in the same spin class, $\partial_{\bar{w}}$ moving it one column to the left and $\partial_{w}$ moving it one column to the right. The operators $\partial_w$ and $\partial_{\bar{w}}$ play the same role as the raising and lowering operators in quantum mechanics and the shift by one in the spin values simply results from the object on which they act being a spin-$1$ director deformation as opposed to a spin-$0$ wavefunction.

The monopole distortions, with $l=0$, result from a rotation of the director by an angle $\theta_0$ in a sphere of radius $a$~\cite{brochard1970theory}. They form a two-real-dimensional vector space for which a basis may be taken to be the distortions $\frac{1}{r}$ and $i \,\frac{1}{r}$. These are shown at the top of Fig.~\ref{fig:3DNematicMultipoles} and can be controllably created in passive nematics using platelet inclusions~\cite{yuan2019elastic}. 

The director distortions of dipole type, with $l=1$, form a six-real-dimensional vector space that splits into two-real-dimensional subspaces for each value of the spin ($0$, $1$, or $2$) as
\begin{align}
    \mathbf{p}^0 & = \biggl\{ \partial_{\bar{w}} \frac{1}{r}, i \,\partial_{\bar{w}} \frac{1}{r} \biggr\} \sim -\frac{1}{2r^3} \bigl\{ x \,\mathbf{e}_x + y \,\mathbf{e}_y , -y \,\mathbf{e}_x + x \,\mathbf{e}_y \bigr\} \sim \frac{1}{r^2} \bigl\{ Y^1_1 , i \,Y^1_1 \bigr\} , \\
    \mathbf{p}^1 & = \biggl\{ \partial_z\frac{1}{r} , i \,\partial_z \frac{1}{r} \biggr\} \sim -\frac{1}{r^3} \bigl\{ z \,\mathbf{e}_x , z\,\mathbf{e}_y \bigr\} \sim \frac{1}{r^2} \bigl\{ Y^0_1 , i \,Y^0_1 \bigr\} , \\
    \mathbf{p}^2 & = \biggl\{ \partial_w \frac{1}{r} , i \,\partial_w \frac{1}{r} \biggr\} \sim -\frac{1}{2r^3} \bigl\{ x \,\mathbf{e}_x - y \,\mathbf{e}_y , y \,\mathbf{e}_x + x \,\mathbf{e}_y \bigr\} \sim \frac{1}{r^2} \bigl\{ Y^{-1}_1 , i \,Y^{-1}_1 \bigr\} .
\end{align}
For comparison, we have presented three representations for the distortions of each spin class: in terms of complex derivatives of $1/r$, two-component vectors whose coefficients are homogenous polynomials of degree $1$ and complex spherical harmonics. In the interest of space we have suppressed certain prefactors in the last of these, but note the difference in sign, and in some cases normalisation, between our representation as complex derivatives and the standard form of the harmonic distortions as two-component vectors \cite{lubensky1998topological}. The two basis functions of any spin class are related by a factor of $i$, which corresponds to a local rotation of the transverse director distortion by $\frac{\pi}{2}$. For a spin-$s$ distortion this is equivalent to a global rotation by $\frac{\pi}{2s}$, with the pair of distortions having the same character and simply providing a basis for all possible orientations. The exception is when $s=0$, such distortions lack an orientation and the local rotation produces two distinct states that transform independently under rotations as a scalar and pseudoscalar. In the dipole case the first is the isotropic distortion recognisable as the UPenn dipole \cite{poulin1997novel} and the second is an axisymmetric chiral distortion with the far-field character of left-handed double twist. Separating $\mathbf{p}^0$ into its isotropic and chiral components allows a decomposition of the dipole director deformations into the basis 
\begin{equation}
    \mathbf{p} = p^I \oplus p^C \oplus \mathbf{p}^1 \oplus \mathbf{p}^2,
\end{equation}
a decomposition which was presented in \cite{pergamenshchik2011dipolar}.

Similarly, the quadrupolar distortions ($l=2$) form a ten-real-dimensional vector space that splits into a sum of two-real-dimensional subspaces for each value of the spin
\begin{align}
    \mathbf{Q}^{-1} & = \biggl\{  \partial^2_{\bar{w}} \frac{1}{r} , i \,\partial^2_{\bar{w}} \frac{1}{r} \biggr\} \sim \frac{3}{4r^5} \bigl\{ (x^2-y^2) \,\mathbf{e}_x + 2xy \,\mathbf{e}_y , -2xy \,\mathbf{e}_x + (x^2-y^2) \,\mathbf{e}_y \bigr\} \sim \frac{1}{r^3} \bigl\{ Y^2_2 , i \,Y^2_2 \bigr\} , \\
    \mathbf{Q}^0 & = \biggl\{ \partial^2_{\bar{w}z} \frac{1}{r} , i \,\partial^2_{\bar{w}z} \frac{1}{r} \biggr\} \sim \frac{3}{2r^5} \bigl\{ xz \,\mathbf{e}_x + yz \,\mathbf{e}_y , -yz \,\mathbf{e}_x + xz \,\mathbf{e}_y \bigr\} \sim \frac{1}{r^3} \bigl\{ Y^1_2 , i \,Y^1_2 \bigr\} , \\
    \mathbf{Q}^1 & = \biggl\{ \partial^2_z \frac{1}{r} , i \,\partial^2_z \frac{1}{r} \biggr\} \sim \frac{1}{r^5} \bigl\{ (2z^2-x^2-y^2) \,\mathbf{e}_x , (2z^2-x^2-y^2) \,\mathbf{e}_y \bigr\} \sim \frac{1}{r^3} \bigl\{ Y^0_2 , i \,Y^0_2 \bigr\} , \\
    \mathbf{Q}^2 & = \biggl\{ \partial^2_{wz} \frac{1}{r} , i \,\partial^2_{wz} \frac{1}{r} \biggr\} \sim \frac{3}{2r^5} \bigl\{ xz \,\mathbf{e}_x - yz \,\mathbf{e}_y , yz \,\mathbf{e}_x + xz \,\mathbf{e}_y \bigr\} \sim \frac{1}{r^3} \bigl\{ Y^{-1}_2 , i \,Y^{-1}_2 \bigr\} , \\
    \mathbf{Q}^3& = \biggl\{ \partial^2_{w} \frac{1}{r} , i \,\partial^2_{w} \frac{1}{r} \biggr\} \sim \frac{3}{4r^5} \bigl\{ (x^2-y^2) \,\mathbf{e}_x - 2xy \,\mathbf{e}_y , 2xy \,\mathbf{e}_x + (x^2-y^2) \,\mathbf{e}_y \bigr\} \sim \frac{1}{r^3} \bigl\{ Y^{-2}_2 , i \,Y^{-2}_2 \bigr\} .
\end{align}
Once again the spin-$0$ distortions can be further partitioned into those that transform as a scalar and pseudoscalar, these being the Saturn's ring distortion \cite{terentjev1995disclination} and a chiral quadrupole with opposing chirality in the two hemispheres, respectively. This yields the basis for the quadrupolar director deformations  
\begin{equation}
    \mathbf{Q} = \mathbf{Q}^{-1} \oplus Q^I \oplus Q^C \oplus \mathbf{Q}^1 \oplus \mathbf{Q}^2 \oplus \mathbf{Q}^3.
\end{equation}

The well-known multipoles, the UPenn dipole and Saturn ring quadrupole, are associated to a configuration of topological defects in the core region and we describe now an extension of this association to all of the multipoles. In general, such an association is not unique, for instance, the colloidal `bubblegum' configuration~\cite{tkalec2009} represents the same far field quadrupole as the Saturn ring, however, for each multipole we can construct a representative arrangement of topological defects which produce it in the far field on the basis of commensurate symmetries and defects of a type and location corresponding to the nodal set of the harmonic. This correspondence allow us to condense the visualisation of complicated three-dimensional fields into a few discrete elements, suggests means by which such distortions might be induced and enables us to build an intuition for their behaviour in active systems through established results for defects~\cite{binysh2020three}.

\begin{figure}[t]
\centering
\includegraphics[width=0.95\textwidth]{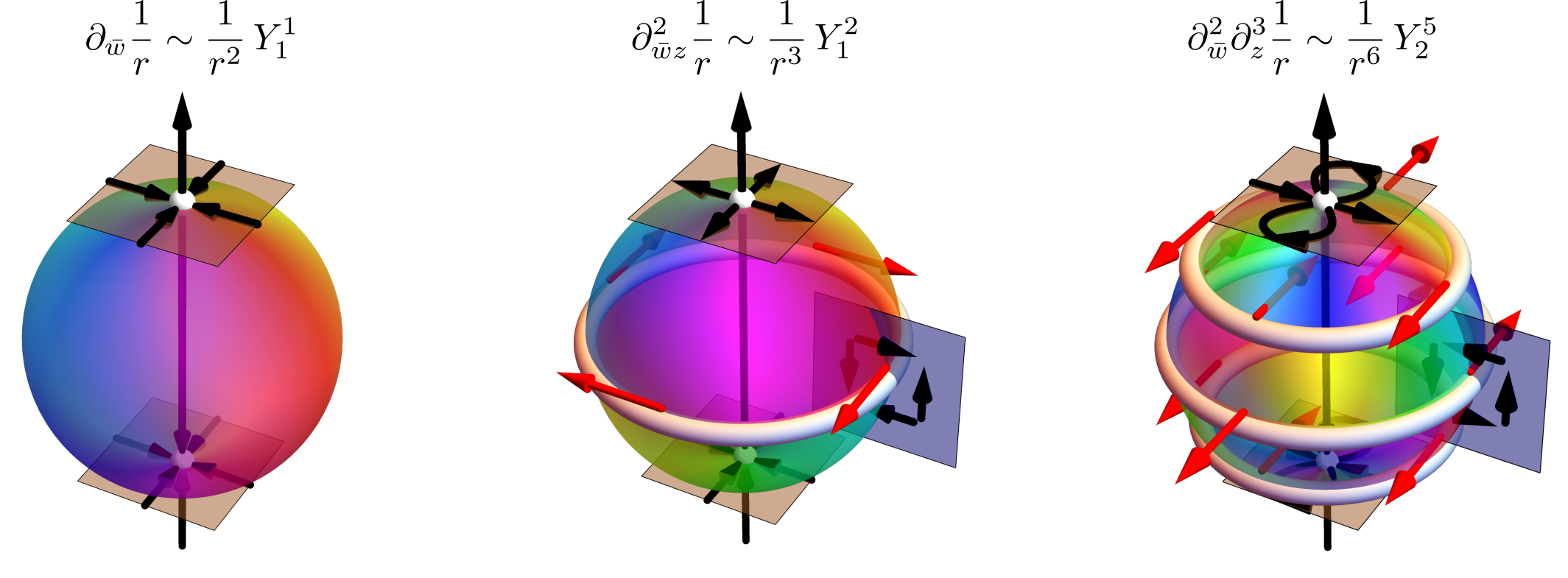}
    \caption[The connection between spherical harmonics and nematic topological defects.]{The connection between spherical harmonics and nematic topological defects. The coloured spheres indicate the phase of the complex spherical harmonics with the nodal set shown in white for simplicity. A representative skeleton of the corresponding nematic distortion is shown in black and the red arrows indicate the winding vector of the director.
    }
    \label{fig:SphericalHarmonicsNematicTopology}
\end{figure}

We first describe some examples, shown in Fig.~\ref{fig:SphericalHarmonicsNematicTopology}. On the left is the spherical harmonic that describes the UPenn dipole, with the form $\partial_{\bar{w}} \frac{1}{r} \sim e^{i\phi}\sin\theta$, visualised on a spherical surface. This has nodes at the two poles about which the phase has $-1$ winding and so we can infer similar winding of the director in the transverse plane. Supplementing with the far-field alignment along $\mathbf{e}_z$ yields the familiar picture of a pair of oppositely charged hedgehog defects.
Similarly, the Saturn ring quadrupole, described by $\partial_{\bar{w}z} \frac{1}{r} \sim e^{i\phi}\sin2\theta$, has zeros at the poles and around the equator. The winding about the poles is still $+1$, but the sign change in the lower hemisphere means that in the transverse plane around the south pole the vector points inwards, resulting in both point defects having topological charge $+1$. With regards to the equatorial line, since the director is everywhere radial the winding vector must be tangential to the defect loop, shown by the red arrows in Fig.~\ref{fig:SphericalHarmonicsNematicTopology}. As the phase changes by $\pi$ on passing from one hemisphere to the other the winding must be $\pm 1$ and the far-field alignment allows us to determine it to be $-1$.
For a general multipole distortion of the form $\partial_{\bar{w}}^m\partial_z^{l-m}(1/r)$ the nodal set is the poles along with $l-m$ lines of latitude. The phase winding of the spherical harmonic dictates the transverse winding of the director and, when supplemented with the far-field alignment, allows us to associate topological point defects with the poles. Similarly, nodal lines may be connected with defect loops with integer winding and a winding vector that rotates according to $e^{im\phi}$. In Fig.~\ref{fig:SphericalHarmonicsNematicTopology} we illustrate this for the case $\partial_{\bar{w}}^2\partial_z^3(1/r)\sim -Y_2^5/r^6$.

We now describe briefly the correspondence for our basis of dipolar and quadrupolar distortions. As already stated, the isotropic scalar in $\mathbf{p}^0$ is the UPenn dipole, its pseudoscalar counterpart a chiral splay-free twist-bend distortion whose integral curves are shown in orange in Fig.~\ref{fig:3DNematicMultipoles}. As a twist-bend mode it may be of particular relevance to extensional systems given their instability to bend distortions. The two dipoles of $\mathbf{p}^1$ are transverse to the far-field alignment, they are related to those resulting from a defect loop of wedge-twist type~\cite{duclos2020topological}. The distortions of $\mathbf{p}^2$ have a hyperbolic character; they describe the far field of a pair of point defects both of which have a hyperbolic structure. Such hyperbolic defect pairs arise in toron configurations in frustrated chiral nematics~\cite{alexander2018topology,smalyukh2010three}.

Similarly, $\mathbf{Q}^0$ contains the Saturn ring quadurpole as the scalar, with the pseudoscalar a pure bend chiral distortion. For the latter, the integral curves of the director possess opposing chirality in the two hemispheres, which could be generated by an appropriately coated Janus particle. The director distortion exhibits a helical perversion in the $z=0$ plane and, being a local rotation of the Saturn ring distortion, may be viewed as resulting from a pair of vortex point defects along with a pure twist defect loop with integer winding. This is similar to the bubblegum defect lines~\cite{tkalec2009,ravnik2009landau} that appear between a colloid diad with normal anchoring, suggesting that this chiral quadrupole could be formed by two colloids with opposing chiral tangential anchoring.

The spin-$1$ quadrupoles consist of pairs of wedge-twist defect loops. The distortions of $\mathbf{Q}^2$ may be associated with a pair of hyperbolic defects along with a defect ring with the appropriate symmetry. The harmonics of spin $-1$ and $3$ contain no $z$-derivatives and so are associated with pairs of point defects only.

\section{Flows from multipole distortions}
\label{sec:multipole_flows}

In this section we calculate the active flow generated by an arbitrary director multipole. We present this initially in vectorial form, converting to the complex representation subsequently. As \eqref{eq:active_Stokes_linearised} is linear the responses due to the two components of $\delta\mathbf{n}$ are independent and so to simplify the derivation we consider only distortions in the $x$-component for now and extend to the general case afterwards. Within this restriction a generic multipole distortion at order $l$ may be written as
\begin{equation}
    \delta n_x = a^{l} \nabla_{{\bf v}_1} \cdots \nabla_{{\bf v}_l}\frac{a}{r} ,
\end{equation}
where ${\bf v}_1, \dots , {\bf v}_l$ are $l$ directions for the differentiation. Substituting this into~\eqref{eq:active_Stokes_linearised} gives the Stokes equation in the form
\begin{equation}
    - \nabla p^{(x)} + \mu \nabla^2 {\bf u}^{(x)} = a^{l+1}\zeta \nabla_{{\bf v}_1} \cdots \nabla_{{\bf v}_l} \Bigl[ {\bf e}_x \,\partial_z + {\bf e}_z \,\partial_x \Bigr] \frac{1}{r} ,
    \label{eq:StokesEquationXComponentInitial}
\end{equation}
where the use of the superscript $^{(x)}$ is to emphasise that we are only treating the response to distortions in the $x$-component of the director.
Taking the divergence of both sides we have
\begin{equation}
- \nabla^2 p^{(x)} + \mu \nabla^2 \nabla\cdot{\bf u}^{(x)} = a^{l+1}\zeta \nabla_{{\bf v}_1} \cdots \nabla_{{\bf v}_l} \partial^2_{xz} \frac{2}{r} .
\end{equation}
Making use of the continuity equation $\nabla \cdot {\bf u}^{(x)} = 0$ in conjunction with the identity $\nabla^2 r = \frac{2}{r}$ we arrive at the solution for the pressure 
\begin{equation}
p^{(x)} = -a^{l+1}\zeta \nabla_{{\bf v}_1} \cdots \nabla_{{\bf v}_l} \,\partial_x \partial_z r = a^{l+1}\zeta \nabla_{{\bf v}_1} \cdots \nabla_{{\bf v}_l} \,\frac{xz}{r^3} .
\end{equation}
Substituting this back into the Stokes equation~\eqref{eq:StokesEquationXComponentInitial} we obtain 
\begin{equation}
    \begin{split}
        \mu \nabla^2 {\bf u}^{(x)} &= a^{l+1}\zeta \nabla_{{\bf v}_1} \cdots \nabla_{{\bf v}_l} \biggl\{ {\bf e}_x \,\partial_z \biggl[ \frac{1}{r} - \partial_x \partial_x r \biggr]- {\bf e}_y \,\partial_x \partial_y \partial_z r + {\bf e}_z \,\partial_x \biggl[ \frac{1}{r} - \partial_z \partial_z r \biggr] \biggr\} ,
    \end{split}
\end{equation}
which can be integrated using the identity $\nabla^2 r^3 = 12 r$ to find
\begin{equation}
        {\bf u}^{(x)} = a^{l+1}\frac{\zeta}{4\mu} \nabla_{{\bf v}_1} \cdots \nabla_{{\bf v}_l}\biggl\{ {\bf e}_x \biggl[ \frac{z}{r} + \frac{x^2 z}{r^3} \biggr] + {\bf e}_y \,\frac{xyz}{r^3} + {\bf e}_z \biggl[ \frac{x}{r} + \frac{xz^2}{r^3} \biggr] \biggr\} .
    \label{eq:FlowGenericxComponent}
\end{equation}

\begin{figure}[t]
\centering
\includegraphics[width=0.96\textwidth]{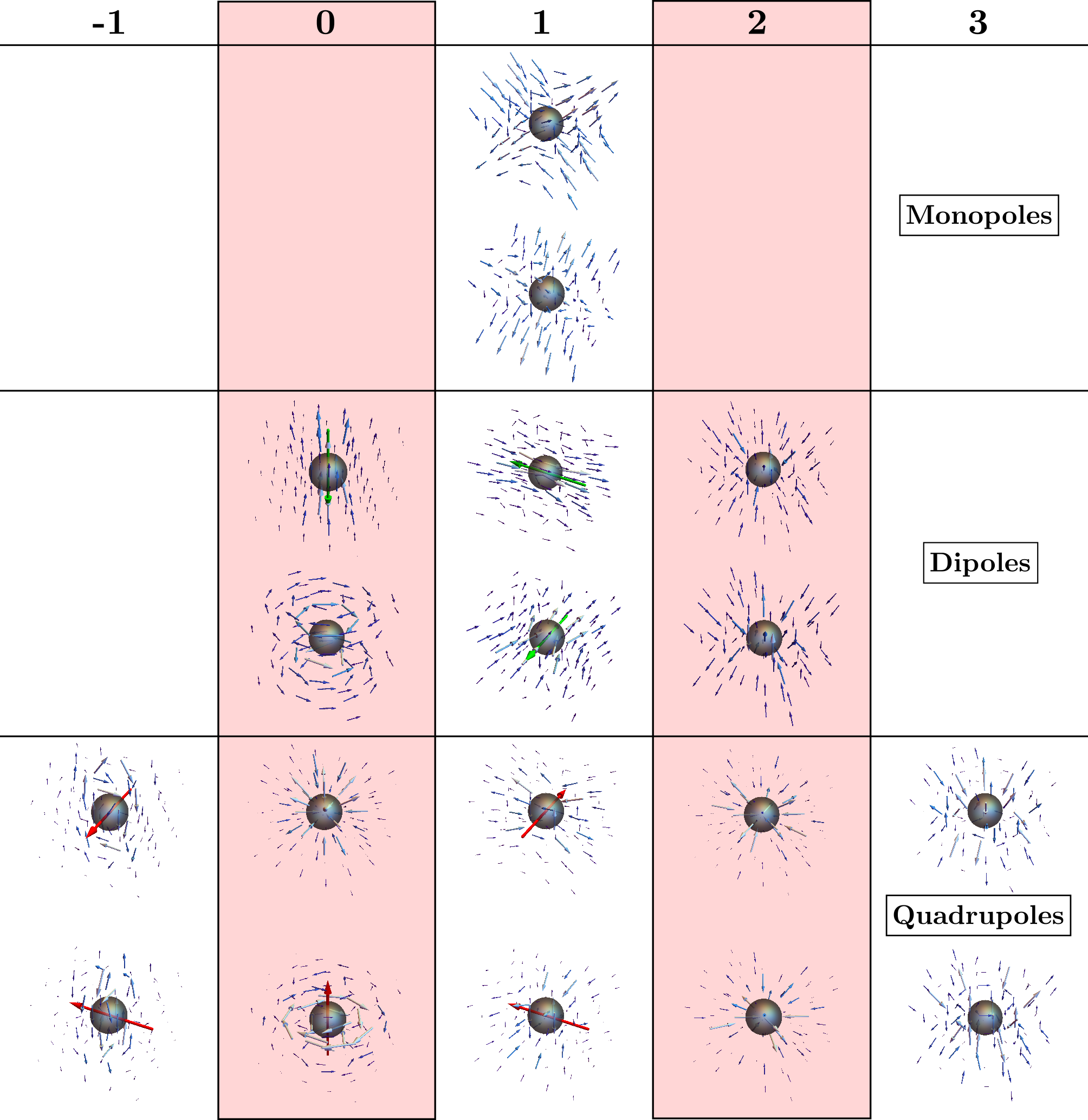}
    \caption[The active flows due to three-dimensional nematic multipole distortions up to quadrupole order.]{The active flows due to three-dimensional nematic multipole distortions up to quadrupole order. The flows are grouped according to their spin, in correspondence with the distortions in Fig.~\ref{fig:3DNematicMultipoles}. Green and red arrows indicate the net active force and torque for the relevant dipoles and quadrupoles respectively, see \S\ref{Chap3:sec:Active forces and torques}.
    }
    \label{fig:3DActiveFlowsNematicMultipoles}
\end{figure}

Both the pressure and flow solutions for a generic multipole distortion are given in terms of derivatives of a fundamental response to a monopole deformation, namely
\begin{gather}
    p^{(x)} = a \zeta \frac{xz}{r^3},
    \label{eq:fundamental_xpressure} \\
    \mathbf{u}^{(x)} = \frac{a\zeta}{4\mu} \left\lbrace {\bf e}_x \biggl[ \frac{z}{r} + \frac{x^2 z}{r^3} \biggr] + {\bf e}_y \,\frac{xyz}{r^3} + {\bf e}_z \biggl[ \frac{x}{r} + \frac{xz^2}{r^3} \biggr] \right\rbrace.
    \label{eq:fundamental_xflow}
\end{gather}
This flow response, shown as the top panel in Fig.~\ref{fig:3DActiveFlowsNematicMultipoles}, is primarily extensional in the $xz$-plane. Interestingly, the flow solution~\eqref{eq:fundamental_xflow} does not decay with distance; this reflects the generic hydrodynamic instability of active nematics~\cite{simha2002hydrodynamic} providing a real-space local response counterpart to the usual Fourier mode analysis. However, the active flow produced by any higher multipole does decay and vanishes at large distances.

The pressure and flow solutions in \eqref{eq:fundamental_xpressure} and \eqref{eq:fundamental_xflow} are complemented by analogous ones resulting from distortions in the $y$-component of the director, obtained by simply interchanging $x$ and $y$. The linearity of~\eqref{eq:active_Stokes_linearised} makes these fundamental responses sufficient to obtain the active flow induced by an arbitrary multipole distortion through taking derivatives appropriate to describe the $x$ and $y$ components of the director, respectively.

We now convert this description to the complex notation used in \S~\ref{Chap3:sec:Multipole director distortions}. This is achieved by taking the combinations $p=p^{(x)}-ip^{(y)}$ and $\mathbf{u}=\mathbf{u}^{(x)}-i\mathbf{u}^{(y)}$. To see this consider the multipole distortion $\delta n=(\mathcal{L}_x+i\mathcal{L}_y)1/r$, where the $\mathcal{L}_i$ are generic real differential operators which generate the $i$-component of the director by acting on $1/r$. This distortion has a conjugate partner given by $i(\mathcal{L}_x+i\mathcal{L}_y)1/r=(-\mathcal{L}_y+i\mathcal{L}_x)1/r$. Acting with this same operator on $\mathbf{u}^{(x)}-i\mathbf{u}^{(y)}$ we have
\begin{equation}
    (\mathcal{L}_x+i\mathcal{L}_y)(\mathbf{u}^{(x)}-i\mathbf{u}^{(y)})=(\mathcal{L}_x\mathbf{u}^{(x)}+\mathcal{L}_y\mathbf{u}^{(y)})-i(-\mathcal{L}_y\mathbf{u}^{(x)}+\mathcal{L}_x\mathbf{u}^{(y)}) ,
\end{equation}
and can see that the flow response for our original distortion forms the real part and that for its conjugate partner the coefficient of $-i$ and the same holds for the pressure response. This leads us to a complex fundamental pressure response
\begin{equation}
    \Tilde{p}=a\zeta\frac{\bar{w}z}{r^3} ,
    \label{eq:FundamentalPressureComplex}
\end{equation}
and, introducing complex basis vectors $\mathbf{e}_{w}=\mathbf{e}_x+i\mathbf{e}_y$ and $\mathbf{e}_{\bar{w}}=\mathbf{e}_x-i\mathbf{e}_y$, a complex-valued fundamental flow vector
\begin{equation}
    \tilde{\mathbf{u}}=\frac{a\zeta}{4\mu}\left\lbrace \mathbf{e}_{w}\, \frac{\bar{w}^2z}{2r^3}+\mathbf{e}_{\bar{w}}\biggl[\frac{z}{r}+\frac{w\bar{w}z}{r^3}\biggr]+\mathbf{e}_z\,\frac{\bar{w}}{r}\left(1+\frac{z^2}{r^2}\right)\right\rbrace.
    \label{eq:FundamentalFlowComplex}
\end{equation}
We use a tilde to distinguish these fundamental responses from those that result due to a generic distortion and which may be found by appropriate differentiation. This provides a unified framework in which the active response to a generic nematic multipole can be calculated through the application of the same complex derivatives that we have used to describe the director distortion. The resulting active flows for distortions up to quadrupole order are shown in Fig.~\ref{fig:3DActiveFlowsNematicMultipoles}, with their layout corresponding to that of the nematic distortions in Fig.~\ref{fig:3DNematicMultipoles} which induce them. 
We now describe some examples in more detail. 

\subsection{UPenn and chiral dipole}
\label{subsec:UPenn_dipole}

\begin{figure}
    \centering
    \includegraphics[width=0.65\textwidth]{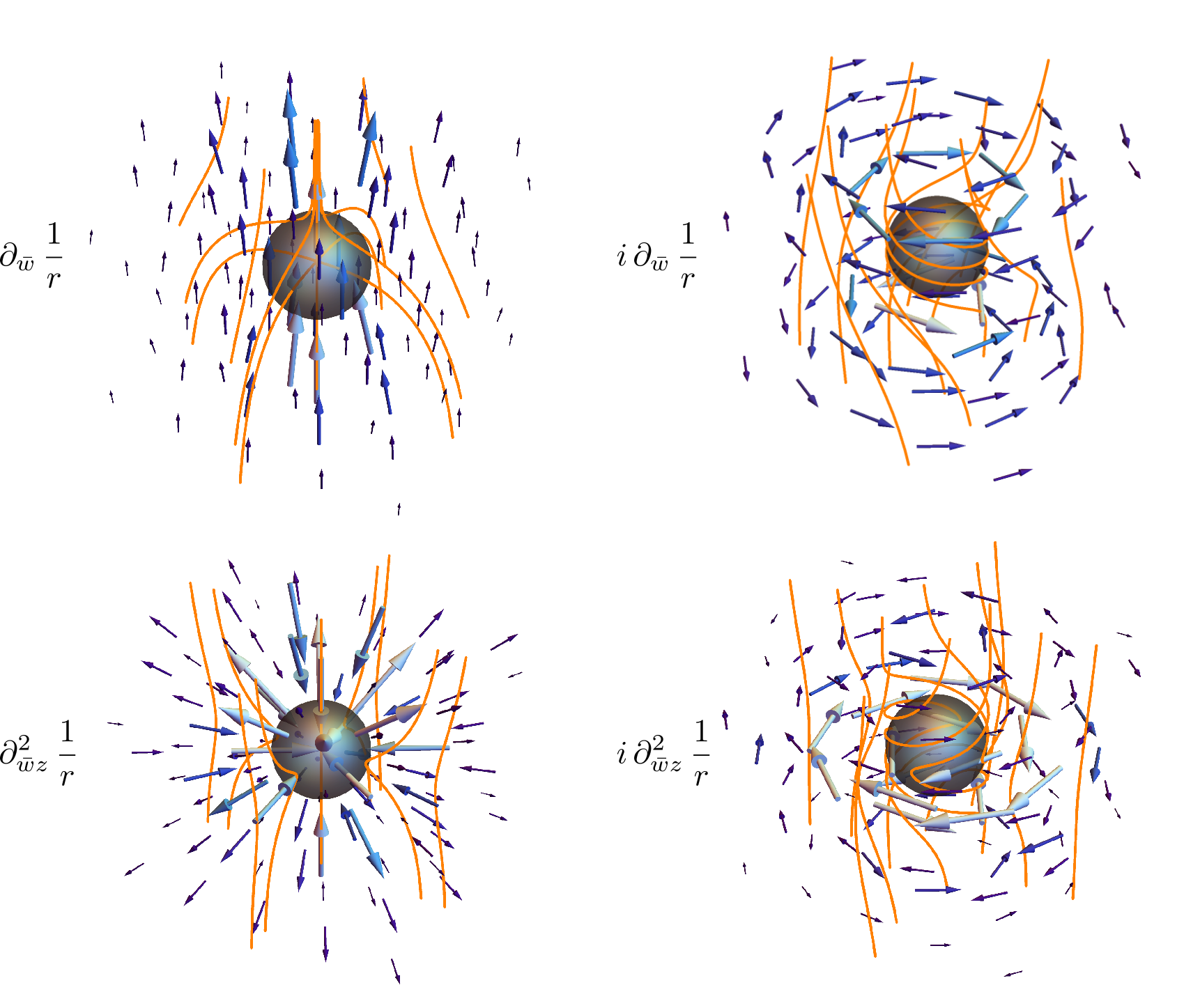}
    \caption[The active flows induced by spin $0$ dipole (top row) and quadrupole (bottom row) distortions.]{The active flows induced by spin $0$ dipole (top row) and quadrupole (bottom row) distortions. The flow is indicated by blue arrows and superposed upon integral curves of the director, shown in orange. On the left are the UPenn dipole and Saturn ring quadrupole and on the right their chiral counterparts.}
    \label{fig:DipoleFlowsHighlight}
\end{figure}

Typically the active responses induced by the two distortions in a spin class will, like the distortions themselves, be related by a global rotation such that while both are needed to form a sufficient basis, the real part essentially serves as a proxy for the pair. This is not true for the spin-$0$ distortions, due to their rotational symmetry, and so we use them in providing an explicit illustration of the active flow calculation. We begin with the UPenn dipole~\cite{poulin1997novel} and its partner the chiral dipole, for which the far-field transverse director is
\begin{equation}
\delta n  \approx  \alpha a \,\partial_{\bar{w}}\frac{a}{r},
\label{eq:dipole_director}
\end{equation}
where $\alpha$ is a dimensionless coefficient, and the corresponding derivative of the fundamental flow solution in \eqref{eq:FundamentalFlowComplex} gives
\begin{equation}
    \begin{split}
        \alpha a \partial_{\bar{w}}\Tilde{\mathbf{u}}=\frac{\zeta\alpha a^2}{4\mu r^5}\left\lbrace\mathbf{e}_{w}\, z\bar{w}(4z^2+w\bar{w})-\mathbf{e}_{\bar{w}}\, 3zw^2\bar{w}+\mathbf{e}_z\, 2\left[3z^4+(z^2+w\bar{w})^2\right]\right\rbrace.
    \end{split}
\end{equation}
Taking the real part gives, after some manipulation, the flow induced by the UPenn dipole as
\begin{equation}
{\bf u} = \alpha a \,\mathfrak{R}\, \partial_{\bar{w}}\Tilde{\mathbf{u}} = \frac{\zeta \alpha a^2}{8\mu} \biggl\{ {\bf e}_z \biggl( \frac{1}{r} + \frac{z^2}{r^3} \biggr) + {\bf e}_r \frac{z}{r^2} \biggl( \frac{3z^2}{r^2} - 1 \biggr) \biggr\} ,
\label{eq:dipole_flow}
\end{equation}
where ${\bf e}_r$ is the unit vector in the radial direction. The flow response to the conjugate distortion, the isotropic chiral dipole is given by
\begin{equation}
{\bf u} = -\alpha a \,\mathfrak{I}\, \partial_{\bar{w}} \Tilde{\mathbf{u}} = -\frac{\zeta \alpha a^2}{4\mu} \frac{z}{r^2}\mathbf{e}_{\phi},
\label{eq:ChiralDipole_Flow}
\end{equation}
with $\mathbf{e}_{\phi}$ the azimuthal unit vector. Both flows decay at large distances like $1/r$ and are highlighted in the top row of Fig.~\ref{fig:DipoleFlowsHighlight}. The UPenn dipole flow has a striking net flow directed along the $z$-axis, reminiscent of that of the Stokeslet flow \cite{blake1974fundamental,chwang1975hydromechanics} associated with a point force along $\mathbf{e}_z$. The chiral dipole generates an axisymmetric flow composed of two counter-rotating vortices aligned along $\mathbf{e}_z$, mirroring the circulating flows produced by spiral defects in two dimensions \cite{khoromskaia2015motility}. The $1/r$ decay of these active vortex flows is unusually slow, slower than the decay of a point torque in Stokesian hydrodynamics \cite{chwang1975hydromechanics}.

Despite the similarity between the active flow induced by the UPenn dipole and a Stokeslet, there is a key difference in their angular dependence. In a Stokeslet, and all related squirming swimmer flows \cite{lighthill1952squirming,pak2014generalized} that result from derivatives of it, the terms with higher angular dependence decay more quickly such that the lowest order terms dominate the far field. By contrast, distortions in active nematics produce asymptotic flow fields in which all terms decay at the same rate regardless of their angular dependence as they all result from the same derivative of the fundamental flow. Thus, even if the same angular terms are present in both systems, the lowest order ones will dominate in the squirming case while the far field will bear the signature of the highest order in the active nematics.

A closer point of comparison comes from the flows induced by active colloids within a passive nematic \cite{daddi2018dynamics,rajabi2020directional}. Calculation of the relevant Green's functions \cite{kos2018elementary} has shown that the anisotropy of the medium leads to a difference in effective viscosities such that a Stokeslet aligned along the director pumps more fluid in this direction. This fits with the anisotropy displayed in \eqref{eq:dipole_flow}, reaffirming the similarity between the flow induced by the UPenn dipole and the Stokeslet.

Considering the pressure response for these distortions in the same way we have
\begin{equation}
    \alpha a \partial_{\bar{w}} \tilde{p} = \frac{\zeta\alpha a^2}{2r^5} z (2z^2-w\bar{w}) = \frac{\zeta\alpha a^2 z}{2r^3} \left(\frac{3z^2}{r^2}-1\right).
\end{equation}
As this expression is purely real it comprises the response due to the UPenn dipole in its entirety; the vanishing of the imaginary part shows that the chiral dipole is compatible with a zero pressure solution. Our complexified construction allows this property to be read off immediately, since $\partial_{\bar{w}}(\bar{w}z^m/r^n)$ will be real for any $m$ and $n$, with this also resulting in the vanishing $z$-component of flow for the chiral dipole. Indeed, this property of pure realness is unchanged by the action of $\partial_z$, it being real itself, and so extends to higher order distortions.

\subsection{Saturn ring and chiral quadrupole}
\label{subsec:Saturn_ring}

Proceeding in the same fashion for the spin-$0$ quadrupoles, for which $\delta n\approx\alpha a^2\partial^2_{\bar{w}z}a/r$, we find that the complexified flow is
\begin{equation}
    \begin{split}
        \alpha a^2\partial^2_{\bar{w}z}\tilde{\mathbf{u}}&=-\frac{\zeta\alpha a^3}{4\mu r^7}\left\lbrace -\mathbf{e}_w\bar{w}(w^2\bar{w}^2+8w\bar{w}z^2-8z^4)+\mathbf{e}_{\bar{w}}3w^2\bar{w}(w\bar{w}-4z^2)\right.\\
        &\left.+\mathbf{e}_z 2z(w^2\bar{w}^2-10w\bar{w}z^2+4z^2)\right\rbrace.
    \end{split}
    \label{eq:Spin0QuadFlowComp}
\end{equation}
Taking the real part gives the flow induced by the Saturn ring quadrupole as
\begin{equation}
    \mathbf{u}=\alpha a^2\mathfrak{R}\partial^2_{\bar{w}z}\tilde{\mathbf{u}}=-\frac{\zeta\alpha a^3}{2\mu r^6}(r^4-12z^2r^2+15z^4)\mathbf{e}_r,
\end{equation}
that is a purely radial flow reminiscent of a stresslet along $\mathbf{e}_z$, shown in the bottom left of Fig.~\ref{fig:DipoleFlowsHighlight}. The purely radial nature is a result of the divergencelessness of the flow, combined with the $1/r^2$ decay and rotational invariance about $\mathbf{e}_z$. Working in spherical coordinates we have
\begin{equation}
    \nabla\cdot\mathbf{u}=\frac{1}{r^2}\partial_r(r^2u_r)+\frac{1}{r\sin\theta}\left[\partial_{\theta}(u_{\theta}\sin\theta)+\partial_{\phi}u_{\phi}\right]=0
\end{equation}
All active flows induced by quadrupole distortions decay as $1/r^2$ and so $\partial_r(r^2u_r)=0$. The distortion is rotationally symmetric and achiral, meaning $u_{\phi}=0$ and the condition of zero divergence reduces to
\begin{equation}
    \frac{1}{r\sin\theta}\partial_{\theta}(u_{\theta}\sin\theta)=0.
\end{equation}
The only non-singular solution is $u_{\theta}=0$, resulting in $u_r$ being the only non-zero flow component. The corresponding pressure is given by
\begin{equation}
    \alpha a^2\partial^2_{\bar{w}z}\tilde{p}=-\frac{3\alpha a^3}{2r^7}(r^4-12z^2r^2+15z^4).
\end{equation}

Taking the imaginary part of~\eqref{eq:Spin0QuadFlowComp} reveals the flow response of the chiral quadrupole to be
\begin{equation}
    \mathbf{u}=-\alpha a^2\mathfrak{I}\partial^2_{\bar{w}z}\tilde{\mathbf{u}}=\frac{\zeta\alpha a^3}{\mu r^2}(3\cos^2\theta-1)\sin\theta\mathbf{e}_{\phi}.
\end{equation}
As illustrated in Fig.~\ref{fig:DipoleFlowsHighlight} this is a purely azimuthal flow corresponding to rotation about the $z$ axis and, as for the chiral dipole, is compatible with a zero pressure solution. The $1/r^2$ decay of this rotational flow is the same as that which results from the rotlet \cite{blake1974fundamental,chwang1975hydromechanics}, but unlike the rotlet the flow direction is not uniform. Rather, as can be seen in Fig.~\ref{fig:DipoleFlowsHighlight}, there is an equatorial band of high-velocity flow accompanied by two slowly counter-rotating polar regions. The distribution of flow speeds is such that the net flow is along $-\mathbf{e}_{\phi}$, consistent with a rotlet along $-\mathbf{e}_z$.

\subsection{Other multipoles}
\label{subsec:other_multipoles}

For the remaining multipoles up to quadrupole order we do not provide the same explicit calculation but instead highlight the key features of the active flows they induce. In full we find that half of the dipole distortions contain directed components in their active flow responses. Along with the isotropic UPenn dipole which produces flow along $\mathbf{e}_z$ the two spin-$1$ dipoles produce directed flows transverse to it. These directed flows indicated that were the source of the distortion free to move it would exhibit active self-propulsion. The net transverse flows for the dipoles of $\mathbf{p}^1$ is in accordance with the previously established motile nature of such defect loops \cite{binysh2020three}. A more complete description of the active dynamics of defect loops via their multipole distortions is presented in Section \ref{subsec:loops_real_structure} and \cite{houston2022defect}.

Along with the chiral dipole, the two additional dipoles which do not generate directed flows are those with spin $2$. These produce active flows which are extensional with the expected two-fold rotational symmetry about the $z$-axis. Direct calculation shows that the flows resulting from spin-$2$ distortions have zero azimuthal component. Once again, this observation is unaffected by $z$-derivatives and so holds true for the higher-order multipoles of the form $\partial^n_z\partial_{w}(1/r)$.

Similarly, there are ten linearly independent quadrupoles, five of which can be seen from Fig.~\ref{fig:3DActiveFlowsNematicMultipoles} to generate rotational flows. As expected, it is the four modes of $\mathbf{Q}^{\pm 1}$ that generate rotations about transverse directions and $Q^C$ that produces rotation around $\mathbf{e}_z$. For two of these, namely those in $\mathbf{Q}^1$, the director distortions are planar, suggesting a two-dimensional analogue and the potential to generate them with cogs or gears \cite{MyThesis}. These distortions may be associated with a pair of opposingly oriented charge-neutral defect loops and so the rotational flow generated by these distortions is in accordance with their antiparallel self-propulsion.

The quadrupoles of $\mathbf{Q}^{-1}$ are composed of pairs of point defects with topological charge $+2$. Using $\partial_{\bar{w}}^2\frac{1}{r}$ as an example, the rotation can be understood by considering the splay distortions in the $xz$ plane. The splay changes sign for positive and negative $x$, leading to antiparallel forces. The active forces are greatest in this plane, as this is where the transverse distortion is radial resulting in splay and bend distortions. Along $\mathbf{e}_y$ the distortions are of twist type and so do not contribute to the active force. This results in the rotational flow shown in Fig.~\ref{fig:3DActiveFlowsNematicMultipoles}. The stretching of the flow along $\mathbf{e}_z$ is as observed for a rotlet in a nematic environment \cite{kos2018elementary}.

Although they lack the rotational symmetry of a stresslet, the flows produced by the quadrupoles of $\mathbf{Q}^2$ are also purely radial. The argument is largely the same as for the Saturn ring distortion, except that the vanishing of $u_{\phi}$ is not due to rotational invariance but a property inherited from the spin-$2$ dipoles.

The quadrupoles of $\mathbf{Q}^3$ produce extensional flows whose spin-$3$ behaviour under rotations about $\mathbf{e}_z$ is commensurate with that of the distortions. Although they visually resemble the similarly extensional flows produced by the dipoles of $\mathbf{p}^2$, they do not share the property of a vanishing azimuthal flow component.

\subsection{Defect loops}
\label{subsec:loops_real_structure}

Of particular relevance to the dynamics of three-dimensional active nematics are charge-neutral defect loops \cite{duclos2020topological,binysh2020three,houston2022defect}. For such defect loops the director field has the planar form 
\begin{equation}
    {\bf n} = \cos \frac{\Upsilon}{4} \,{\bf e}_z + \sin \frac{\Upsilon}{4} \,{\bf e}_x ,
    \label{eq:director_defect_loop}
\end{equation}
where $\Upsilon$ is the solid angle function for the loop~\cite{maxwell1873treatise,binysh2018maxwell}, and is a critical point of the Frank free energy in the one-elastic-constant approximation~\cite{friedel1969boucles}. This allows a multipole expansion for the director at distances larger than the loop size in which the multipole coefficients are determined explicitly by the loop geometry~\cite{houston2022defect} 
\begin{equation}
    \Upsilon(\mathbf{x}) = \frac{1}{2} \int_K \epsilon_{ijk} \,y_j \,\text{d}y_k \,\partial_i\frac{1}{r} - \frac{1}{3} \int_K \epsilon_{ikl} \,y_l y_k \,\text{d}y_l \,\partial_i\partial_j \frac{1}{r} + \dots ,
    \label{eq:omega_multipole}
\end{equation}
where $\mathbf{y}$ labels the points of the loop $K$ and $r=|\mathbf{x}|$ with the `centre of mass' of the loop defined to be at $\mathbf{x}=0$. The dipole moment vector is the projected area of the loop, while the quadrupole moment is a traceless and symmetric tensor with an interpretation via the first moment of area or, in the case of loops weakly perturbed from circular, the torsion of the curve.

The planar form of the director field~\eqref{eq:director_defect_loop} corresponds to a restricted class of director deformations in which $\delta n$ is purely real. This disrupts the complex basis we have adopted for the representation of multipoles, so that another choice is to be preferred. We may say that the planar director selects a real structure for the orthogonal plane $\mathbb{C}$, breaking the $U(1)$ symmetry, and the restricted multipoles should then be decomposed with respect to this real structure. Accordingly, the pressure and flow responses may be generated by derivatives of the fundamental responses for distortions in $\mathbf{e}_x$, \eqref{eq:fundamental_xpressure} and \eqref{eq:fundamental_xflow}, with these derivatives corresponding to the multipole expansion of the solid angle shown in \eqref{eq:omega_multipole}. The details of this approach along with the consequences it has for both the self-propulsive and self-rotational dynamics of active nematic defect loops are given in \cite{houston2022defect}.

\subsection{Technical note}
\label{subsec:technical_note}

\begin{figure}[t]
    \centering
    \includegraphics[width=0.65\textwidth]{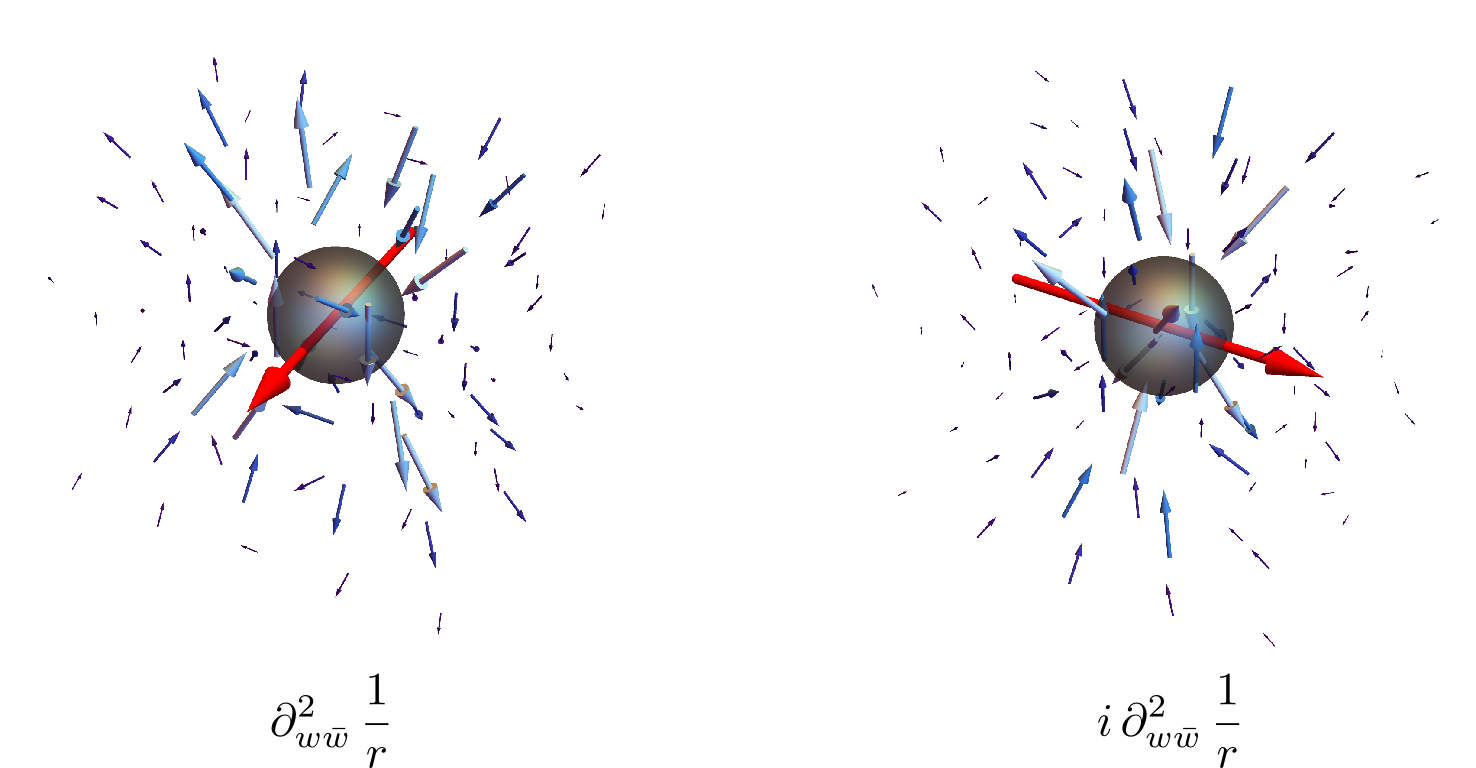}
    \caption[Additonal flow solutions induced by spin-1 nematic multipoles.]{Additonal flow solutions induced by spin-1 nematic multipoles. The nematic multipoles which induce the flows are shown below them as complex derivatives of $1/r$. The red arrows indicate the net active torque.}
    \label{fig:3DActiveFlowsNematicMultipolesAdditional}
\end{figure}

We conclude this section with a technical note on the flow solutions that we have presented. The construction for calculating active flow responses that we have developed in this section requires knowledge of the multipole as a specified set of derivatives of $1/r$. The harmonic director components satisfy $\nabla^2n_i \propto \delta(\mathbf{r})$ and while this delta function does not affect the far-field director it impacts the flow solutions. 
Consequently, at quadrupole order and higher, distinct derivatives of $\frac{1}{r}$ can produce the same multipole distortion in the director but have different associated active flows. As an explicit example we take the spin-$1$ quadrupole shown in Fig.~\ref{fig:3DNematicMultipoles}, which may be written as $\mathbf{n}=a^2\partial_z^2 \frac{a}{r} \,\mathbf{e}_x + \mathbf{e}_z$ and therefore induces an active flow given by the action of $a^2\partial_z^2$ on \ref{eq:FundamentalFlowComplex}, as is illustrated in Fig.~\ref{fig:3DActiveFlowsNematicMultipoles}. However the same director distortion is captured by $\mathbf{n} = -4 a^2\partial_{w\bar{w}}^2 \frac{a}{r} \,\mathbf{e}_x + \mathbf{e}_z$, for which the corresponding active flow is shown in Fig.~\ref{fig:3DActiveFlowsNematicMultipolesAdditional}. A partial resolution to this ambiguity is that any non-equilibrium phenomenological features such as propulsion or rotation will be invariant to this choice of derivatives since, as we shall show in the following section, they can be expressed directly in terms of the director components. As a more complete resolution we reiterate that whenever an exact solution for the director is known the appropriate derivatives can be determined, as demonstrated earlier for defect loops \cite{houston2022defect}, and so the apparent ambiguity disappears.

\section{Active forces and torques}
\label{Chap3:sec:Active forces and torques}
The directed and rotational active flow components highlighted above result in viscous stresses whose net effect must be balanced by their active counterparts, since the net force and torque must be zero. Consequently, these generic aspects of the response of an active nematic can be identified by considering the contribution that the active stresses make to the force and torque
    \begin{gather}
        \mathbf{f}^a = \int \zeta {\bf nn} \cdot \text{d}{\bf A} \approx \int \zeta \biggl\{ {\bf e}_x \frac{z \,\delta n_x}{r} + {\bf e}_y \frac{z \,\delta n_y}{r} + {\bf e}_z \frac{x \,\delta n_x + y \,\delta n_y}{r} \biggr\} \text{d}A , 
        \label{eq:active_force} \\[2mm]
        \begin{split}
            \boldsymbol{\tau}^a = \int {\bf x} \times \zeta {\bf nn} \cdot \text{d}{\bf A} & \approx \int \zeta \biggl\{ {\bf e}_x \biggl[ \frac{xy \,\delta n_x}{r} + \frac{(y^2-z^2) \delta n_y}{r} \biggr] + {\bf e}_y \biggl[ \frac{(z^2-x^2) \delta n_x}{r} - \frac{xy \,\delta n_y}{r} \biggr] \\
            & \qquad + {\bf e}_z \frac{z(-y \,\delta n_x + x \,\delta n_y)}{r} \biggr\} \text{d}A , 
        \end{split} \label{eq:active_torque}
    \end{gather}
integrating over a large sphere of radius $r$. These integrals depend on the surface of integration, as the active stresses are neither divergenceless nor compactly supported. However, a spherical surface is concordant with the multipole approach we are taking and the results are then independent of the radius, as a direct consequence of the orthogonality of spherical harmonics. From these expressions we can read off the multipole that will generate any desired active force or torque; dipoles generate forces and quadrupoles generate torques. When the active torque is non-zero, the compensating viscous torque will drive a persistent rotation of the multipole, creating an active ratchet; similarly, a non-zero active force will generate directed fluid flow. The above integrals therefore provide a solution to the inverse problem: given a particular non-equilibrium response, which distortion induces it? Hence they serve as a design guide for generating out of equilibrium responses in active nematics.

If the multipole is free to move it will self-propel and rotate. The translational and rotational velocities are related to the viscous forces and torques by a general mobility matrix~\cite{kim1991microhydrodynamics}. 
In passive nematics, experiments~\cite{loudet2004stokes} and simulations~\cite{ruhwandl1996friction,stark2001stokes} have found that it is sufficient to take a diagonal form for the mobility (no translation-rotation coupling) with separate viscosities for motion parallel, $\mu_{\parallel}$, and perpendicular, $\mu_{\perp}$, to the director, with typical ratio of viscosities $\mu_{\perp}/\mu_{\parallel} \sim 1.6$ \cite{loudet2004stokes,ruhwandl1996friction,stark2001stokes}. This has the consequence that in general the force and velocity are not colinear 
\begin{equation}
    {\bf U} = \frac{-1}{6\pi a} \biggl[ \frac{1}{\mu_{\parallel}} f^{a}_{\parallel} \,{\bf e}_z + \frac{1}{\mu_{\perp}} \,{\bf f}^{a}_{\perp} \biggr] .
\end{equation}
We again use the UPenn dipole as an example. Integrating the active stresses over a spherical surface of radius $R$ we find an active force 
\begin{equation}
\int \zeta {\bf nn} \cdot \text{d}{\bf A} \approx -\frac{\zeta \alpha a^2}{2} \int \biggl\{ {\bf e}_x \frac{xz}{R^4} + {\bf e}_y \frac{yz}{R^4} + {\bf e}_z \biggl[ \frac{z}{R} + \frac{x^2+y^2}{R^4} \biggr] \biggr\} \text{d}A = -\frac{4\pi \zeta \alpha a^2}{3} \,{\bf e}_z .
\end{equation}
Balancing this against Stokes drag predicts a `self-propulsion' velocity for the active dipole of 
\begin{equation}
{\bf U} = \frac{2\zeta \alpha a}{9\mu_{\parallel}} \,{\bf e}_z .
\label{eq:dipole_propulsion}
\end{equation}
For extensile activity ($\zeta > 0$) the dipole moves `hyperbolic hedgehog first' and with a speed that increases linearly with the core size $a$. This self-propulsion is in accordance with the directed component of the active flow, as can be seen in Fig.~\ref{fig:DipoleFlowsHighlight}. The same self-propulsion speed along $\mathbf{e}_x$ and $\mathbf{e}_y$ is found for the transverse dipoles of $\mathbf{p}^1$, except that the parallel viscosity $\mu_{\parallel}$ should be replaced with $\mu_{\perp}$. Again, this self-propulsion agrees with the directed flow induced by these distortions, as calculated through the multipole approach, shown in Fig.~\ref{fig:3DActiveFlowsNematicMultipoles} \cite{houston2022defect} and also with the results of both a local flow analysis and simulations \cite{binysh2020three}. The same directed motion has been observed in a related system of an active droplet within a passive nematic \cite{rajabi2020directional}, with the droplet inducing a UPenn dipole in the nematic and moving in the direction of the hedgehog defect at a speed that grew with the droplet radius. The mechanism at play is different however; the motion results from directional differences in viscosity resulting from the anisotropic environment.

To illustrate the rotational behaviour we use a member of $\mathbf{Q}^1$, $\partial^2_z(1/r)$, as an example. We find an active torque
\begin{align}
    \int \zeta\mathbf{x}\times\mathbf{n}\mathbf{n}\cdot \text{d}\mathbf{A}&\approx\zeta\alpha a^3\int\frac{1}{r^6}(2z^2-x^2-y^2)\left\lbrace xy\mathbf{e}_x+(z^2-x^2)\mathbf{e}_y-yz\mathbf{e}_z \right\rbrace \text{d}A\\
    & =\frac{8\pi\zeta\alpha a^3}{5}\mathbf{e}_y.
    \label{eq:ActiveTorqueExample}
\end{align}
Balancing against Stokes drag as was done in the dipole case gives an angular velocity
\begin{equation}
    \mathbf{\Omega}=-\frac{\zeta\alpha}{5\mu}\mathbf{e}_y.
\end{equation}
We note that for this and all other distortions which result in net torques the angular velocity is independent of the colloid size. In accordance with the relation $\partial^2_z+4\partial^2_{w\bar{w}}(1/r)=0$, the torque resulting from $\partial^2_{w\bar{w}}(1/r)$ is of the opposite sign and a quarter the strength. The net active torques due to harmonics of $\mathbf{Q}^0$ and $\mathbf{Q}^{-1}$ have the directions indicated in Fig.~\ref{fig:3DActiveFlowsNematicMultipoles} and half the magnitude of \eqref{eq:ActiveTorqueExample}.

Let us consider the approximate magnitude of the effects we have described. Beginning with the self-propulsion speed, the fluid viscosity is roughly $10^{-2}$~Pa s \cite{shendruk2017dancing}, although effects due to the elongated form of the nematogens could increase this by a factor of $30$ or so \cite{batchelor1971stress,lang2019effects}. Both the activity \cite{giomi2014defect} and the dipole moment constant \cite{lubensky1998topological} are of order unity, meaning the colloid would approximately cover its radius in a second. Similar approximations for the quadrupole give an angular velocity of about $2/3$~rad s$^{-1}$. For a colloid of radius $10~\mu$m this has an associated power of the order of femtowatts, the same as predicted for bacterial ratchets \cite{sokolov2010swimming}.

\section{Two-dimensional systems and ratchets}
As noted above, the planar nature of the rotational distortions in $\mathbf{Q}^1$ suggests the existence of two-dimensional analogues. In part motivated by this we now discuss the active response of multipolar distortions in two dimensions, again beginning with the connection between these multipoles and topological defect configurations.

\subsection{Multipoles and topological defects}
The categorisation of the harmonic distortions in two dimensions is much simpler, but we provide it here for completeness. Taking the asymptotic alignment to be along $\mathbf{e}_y$ the symmetry of the far-field director is now described by the order $2$ group $\left\lbrace 1,R_y \right\rbrace$, with $R_y$ reflection with axis $\mathbf{e}_y$, under which the monopole distortion $n_x\sim A\log(r/a)$ is antisymmetric. The higher-order distortions are once again generated via differentiation of the monopole, with $\partial_y$ leaving the symmetry under $R_y$ unchanged and $\partial_x$ inverting it. 

It should be noted that the potential multiplicity of differential representations of harmonics that arose in three dimensions does not occur in two dimensions. This is because, under the assumption of a single elastic constant, the director angle $\phi$ may be written as the imaginary part of a meromorphic function of a single complex variable and this naturally defines the appropriate set of derivatives. Making $z=x+iy$ our complex variable we write $\phi=\mathfrak{I}\left\lbrace\mathfrak{f}(z)\right\rbrace$ which upon performing a Laurent expansion of $\mathfrak{f}(z)$ around $z=0$ and assuming the existence of a uniform far-field alignment gives
\begin{equation}
        \phi = \mathfrak{I}\left\lbrace\sum_{n=-\infty}^{0}a_nz^n\right\rbrace = \mathfrak{I}\left\lbrace a_0+\sum_{n=1}^{\infty}(-1)^{n-1}\frac{a_n}{(n-1)!}\partial^n_z(\ln z)\right\rbrace.
\end{equation}
Hence at every order there is a one parameter family of distortions, corresponding to the phase of the $a_n$. A natural basis at order $n$ is provided by $\left\lbrace \mathfrak{R}\left\lbrace\partial^n_z(\ln z)\right\rbrace,\mathfrak{I}\left\lbrace\partial^n_z(\ln z)\right\rbrace \right\rbrace$. This basis consists of a symmetric and anti-symmetric distortion under the action of $R_y$, the roles alternating with order, and of course correspond to the two harmonic functions $\cos n\theta/r^n$ and $\sin n\theta/r^n$.

In two dimensions the connection between defect configurations and far-field multipole distortions can be made concrete, and also serves as an illustration of how a particular set of derivatives is determined. For defects with topological charges $s_j$ at locations $z_j$ the angle that the director makes to $\mathbf{e}_x$ is given by
\begin{equation}
\begin{split}
    \phi=\phi_0+\sum_{j}s_j\mathfrak{I}\left\lbrace\ln\left(\frac{z-z_j}{a}\right)\right\rbrace,\\
\end{split}
\label{eq:DirectorAngle2DExact}
\end{equation}
which, upon performing a series expansion, gives
\begin{align}        \phi&=\phi_0+\sum_js_j\mathfrak{I}\left\lbrace\ln(z/a)\right\rbrace-\sum_{n=1}^{\infty}\frac{\mathfrak{I}\left\lbrace\sum_js_jz_j^n\Bar{z}^n\right\rbrace}{n|z|^{2n}} ,\\
\begin{split}       &=\phi_0+\sum_js_j\mathfrak{I}\left\lbrace\ln(z/a)\right\rbrace+\sum_{n=1}^{\infty}\frac{(-1)^n\mathfrak{I}\left\lbrace\sum_js_jz_j^n\partial^n_z\ln z\right\rbrace}{n!} ,
\end{split}
\label{eq:DirectorAngle2DSeriesGeneric}
\end{align}
Provided the total topological charge is zero the winding term proportional to $\ln w$ vanishes and $\phi_0$ is the far-field alignment. The distortions are given as a series of harmonics in which the coefficient of the $n^{\text{th}}$ harmonic is determined by a sum of $z_j^n$ weighted by the defect charges.

We would like to have a basis of representative defect configurations for each harmonic distortion. However, it can be seen from \eqref{eq:DirectorAngle2DSeriesGeneric} that the correspondence between arrangements of topological defects and the leading order nematic multipole is not one-to-one. Two defect-based representations of harmonic will prove particularly useful to us. The first, which we develop in this chapter, provides a representation in terms of half-integer defects on the disc and allows an intuition for the response to multipole distortions in active nematics through known results for such defects \cite{giomi2013defect,giomi2014defect}. The second uses the method of images to construct defect arrangements corresponding to a specific anchoring condition on the disc, with the same multipoles dominating the nematic distortion in the far field. This representation naturally lends itself to the control of induced multipoles through colloidal geometry and is explored fully in \cite{MyThesis}. Nonetheless, both of these representations will be of use to us in the remainder of this chapter and as they are equally valid near-field representations for the asymptotic distortions that we are considering we will pass fairly freely between them.

With this aforementioned half-integer representation in mind, let us consider sets of $2m$ defects sitting on the unit circle, with $-1/2$ defects at the $m^{\text{th}}$ roots of unity and $+1/2$ defects at the intermediate points. A useful formula here is the following for the sum of a given power of these roots of unity, after first rotating them all by a given angle $\theta$
\begin{equation}
    \sum_{k=0}^{m-1}\left(e^{i\theta}e^{i\frac{2\pi}{m}k}\right)^n=
    \begin{cases}
    me^{in\theta},& \text{if } m|n \\ 0, & \text{otherwise}
    \end{cases}.
\end{equation}
The vanishing of this sum for values of $n$ that are not multiples of $m$ comes directly from the expression for the geometric sum and is a consequence of the cyclic group structure of the roots of unity. It means that the lowest order multipole distortion induced by such an arrangement of defects is order $m$ and so allows a desired multipole distortion to be selected as the dominant far-field contribution. Explicitly, the director angle is given by
\begin{equation}
    \phi=\phi_0+\sum_{k\text{ odd}}\frac{\mathfrak{I}\left\lbrace\Bar{z}^{mk}\right\rbrace}{k|z|^{2mk}}
    =\phi_0+\frac{\mathfrak{I}\left\lbrace\Bar{z}^{m}\right\rbrace}{|z|^{2m}}+O\left(\frac{1}{z^{3m}}\right),
\end{equation}
with the approximation becoming rapidly better for higher-order multipoles due to the condition that $n$ must be an odd multiple of the number of defects. Rotating the entire set of defects rigidly by an angle $-\pi/(2m)$ generates the conjugate multipole as the dominant far-field contribution
\begin{equation}
    \phi=\phi_0+\sum_{k\text{ odd}}\frac{\mathfrak{I}\left\lbrace (-i)^k\Bar{z}^{mk}\right\rbrace}{k|z|^{2mk}}
    =\phi_0-\frac{\mathfrak{R}\left\lbrace \Bar{z}^{m}\right\rbrace}{|z|^{2m}}+O\left(\frac{1}{z^{3m}}\right),
\end{equation}
with the natural interpolation between these two harmonics as the defect configuration is rigidly rotated.

\begin{figure}
    \centering
    \includegraphics[width=0.6\textwidth]{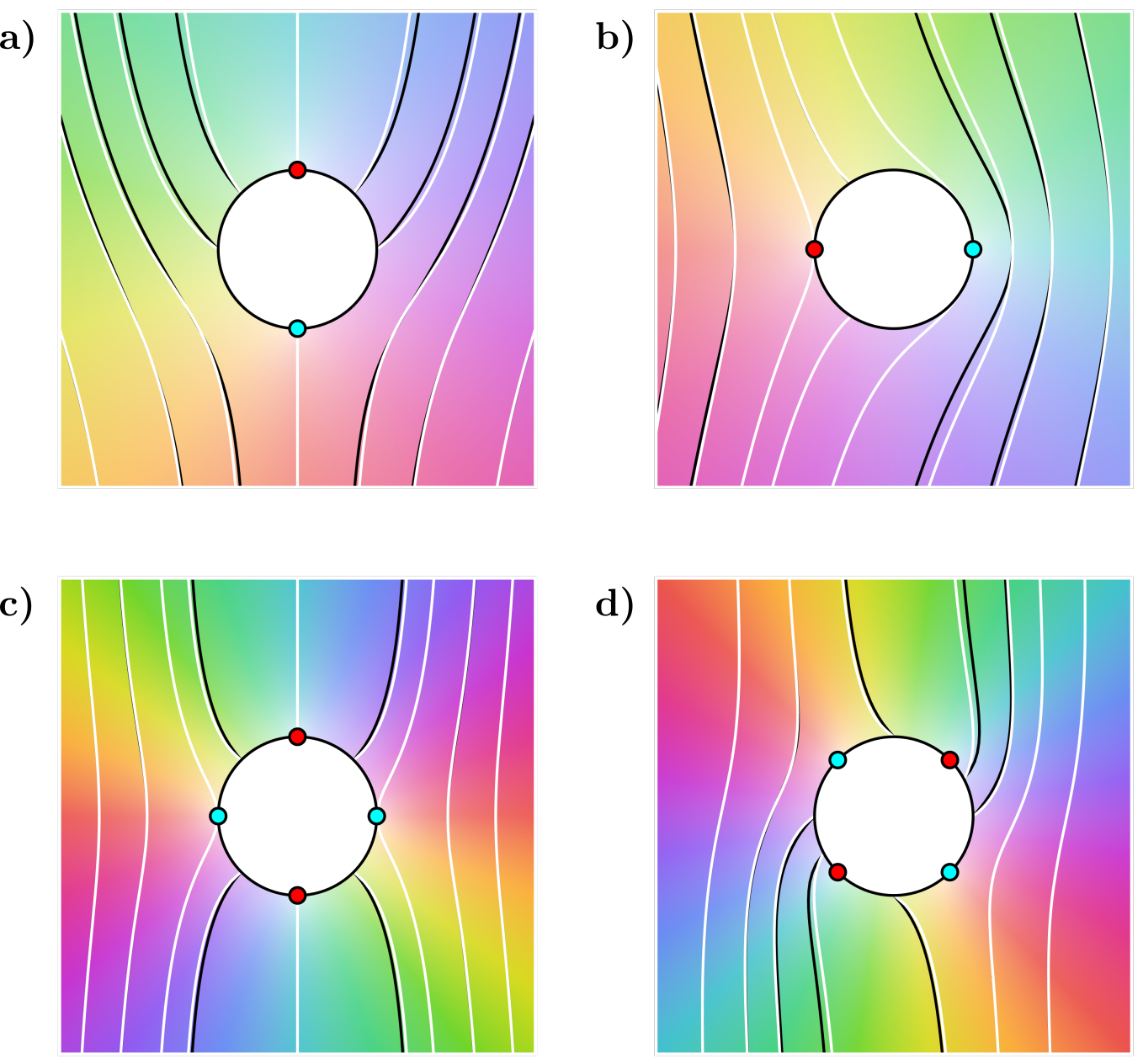}
    \caption[Representative defect configurations for nematic multipoles in two dimensions.]{Representative defect configurations for nematic multipoles in two dimensions. The red and cyan dots indicate the locations of $+1/2$ and $-1/2$ defects respectively. The black curves are the integral curves of the corresponding director field and the background colour shows the phase of the complex function whose imaginary part gives the exact director angle, as in \eqref{eq:DirectorAngle2DExact}. The white lines are the integral curves of the dominant multipole, that is the leading term of \eqref{eq:DirectorAngle2DSeriesGeneric}. The multipole series converges onto the exact director angle outside a core region, shown as a white disc, and the leading multipole provides a remarkably good approximation in this region. 
    }
    \label{fig:2DMultipolesDefects}
\end{figure}

Hence we can interchange between a given harmonic distortion and a defect arrangement which has this harmonic as its dominant far-field contribution, with the correspondence becoming rapidly more accurate for higher orders, allowing us to relate the existing results for the behaviour of active defects \cite{giomi2013defect,giomi2014defect} to ours and vice versa. This correspondence is illustrated in Fig.~\ref{fig:2DMultipolesDefects}. The locations of $+1/2$ and $-1/2$ defects are indicated with red and cyan dots respectively and the background colouring denotes the phase of the complex function $\sum s_j\ln(z-z_j)$, whose imaginary part provides the director angle for the given defect arrangement. The integral curves of this director field are shown in black and are remarkably well matched by those of the leading multipole, shown in white, despite the asymptotic nature of the approximation. In this context we are able to make precise the notion of a core region of a singular distortion, outside of which our multipole approach applies. The series in \eqref{eq:DirectorAngle2DSeriesGeneric} is attained through a Taylor series of terms of the form $\ln(1-1/z)$, which are convergent for $|z|>1$. More generally the greatest radial displacement of a defect defines a core radius, outside of which the multipole series converges onto the exact director angle.

\subsection{Flows from multipole distortions}
\label{Chap3:subsec:Flows from multipole distortions}
We can proceed analogously to our three-dimensional calculation in generating the active flows from a fundamental response in two dimensions, provided we are mindful of the logarithmic form that the monopole now has. A director rotation by $\theta_0$ inside a disc of radius $a$ results in an equilibrium texture given by
\begin{equation}
    \mathbf{n}=\cos\left(\frac{\theta_0\log(r/R)}{\log(a/R)}\right)\mathbf{e}_y+\sin\left(\frac{\theta_0\log(r/R)}{\log(a/R)}\right)\mathbf{e}_x,
\end{equation}
which in the far field tends to a monopole distortion $\mathbf{n}\approx\mathbf{e}_y+\frac{\theta_0\log(r/R)}{\log(a/R)}\mathbf{e}_x$. Due to the logarithmic divergence of the fundamental harmonic in two dimensions it is necessary to normalise through a large length $R$ such that a uniformly aligned far-field director is recovered.

Following our three-dimensional analysis we solve Stokes' equations to linear order in nematic deformations for a monopole distortion. We write Stokes' equations in terms of complex derivatives as
\begin{equation}
    2\partial_{\bar{z}}(-p+i\mu\omega)=f,
\end{equation}
where we have used that $2\partial_z u=\nabla\cdot\mathbf{u}+i\omega$, with $\omega$ the vorticity. Hence we seek $f$ as a $\bar{z}$-derivative, implicitly performing a Helmholtz derivative with the real and imaginary parts of the differentiated term corresponding to the scalar and vector potentials respectively. Expressing the active force in this way we have
\begin{equation}
    2\partial_{\bar{z}}(-p+i\mu\omega)=\frac{\zeta\theta_0}{\log(a/R)}\partial_{\bar{z}}\left(\frac{i\bar{z}}{z}\right)
\end{equation}
and so
\begin{equation}
    -p+i\mu\omega=\frac{\zeta\theta_0}{2\log(a/R)}\frac{i\bar{z}}{z}.
    \label{eq:StokesComplexAchiral}
\end{equation}
Reading off the pressure and vorticity, solving for the flow and converting back to Cartesians the fundamental flow response is now found to be
\begin{gather}
    \Tilde{{\bf u}} = \frac{\zeta\theta_0}{8\mu\log(a/R)} \biggl[\frac{x^2-y^2}{r^2}(-y\mathbf{e}_x+x\mathbf{e}_y)+ 2\log\left(\frac{r}{R}\right)(y\mathbf{e}_x+x\mathbf{e}_y)\biggr], 
    \label{eq:active_flow_2D} \\[2mm]
    \begin{split}
    \Tilde{p} = -\frac{\zeta \theta_0}{\log(a/R)} \frac{xy}{r^2}.
    \label{eq:active_pressure_2D}
    \end{split}
\end{gather}
There is a clear similarity between these solutions and their three-dimensional counterparts, but while the fundamental flow response is still extensional it now grows linearly with distance from the distortion, with this change in scaling inherited by the subsequent harmonics.

As in the three-dimensional case we can gain general insight into the active response of a nematic by considering the net contribution of the active stresses to the force and torque when integrated over a large circle of radius $r$
\begin{gather}
    \int\zeta\mathbf{n}\mathbf{n}\cdot\mathbf{e}_r \text{d}r\approx\int\zeta\left\lbrace\frac{y\delta n_x}{r}\mathbf{e}_x+\frac{x\delta n_x}{r}\mathbf{e}_y\right\rbrace \text{d}r, \\    \int\mathbf{x}\times\zeta\mathbf{n}\mathbf{n}\cdot\mathbf{e}_r\text{d}r\approx\int\zeta\frac{(y^2-x^2)\delta n_x}{r}\text{d}r.
\end{gather}
We see that in two dimensions both dipoles will self-propel if free to move and there is a single chiral quadrupole which produces rotations.

\begin{figure}
    \centering
    \includegraphics[width=0.55\textwidth]{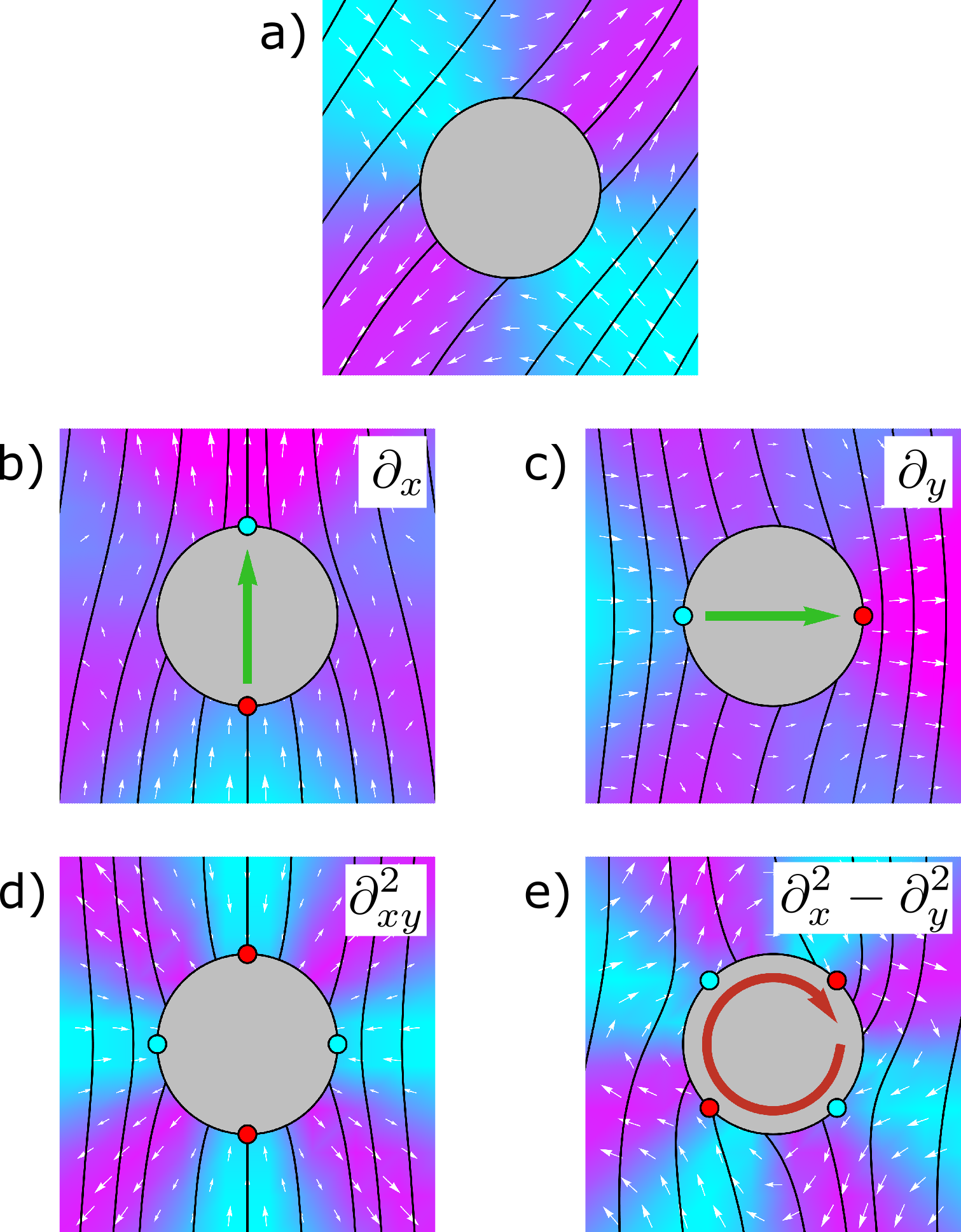}
    \caption[Distortions up to quadrupole order in two-dimensional active nematics.]{Distortions up to quadrupole order in two-dimensional active nematics. The active flow in white is superposed on the pressure field, with the integral curves of the director shown in black. (a) The fundamental monopole response is extensional and grows linearly with distance from the distortion. (b) and (c) show the flows induced by dipole distortions, labelled by the appropriate derivative of the nematic monopole, with the green arrows indicating the direction of self-propulsion that would result from net active forces in extensile systems. The vertical and horizontal dipoles are the far-field director responses to normal and tangential anchoring respectively and may also be interpreted as arising from a pair of $+1/2$ (cyan) and $-1/2$ (red) defects. The self-propulsion matches that expected for the $+1/2$ defect.
    }
    \label{fig:2DActiveMonoDi}
\end{figure}

The far-field flow solutions for distortions up to dipole order are illustrated in Fig.~\ref{fig:2DActiveMonoDi}, superposed over the nematic director. Both dipoles are now motile and as in the three-dimensional case they set up flows reminiscent of the Stokeslet. Vertical and horizontal self-propulsive modes may be viewed as resulting from normal and tangential anchoring respectively of the nematic on a disc. Interpolating between these orthogonal modes the angle of motility changes commensurately with the anchoring angle, such that sufficient control of the boundary conditions would allow for self-propulsion at an arbitrary angle with respect to the far-field alignment. This change in the dipole character can be represented by rigidly rotating the defect pair around the unit circle and the resulting motility is as would be expected from the position and orientation of the $+1/2$ defect \cite{giomi2014defect,vromans2016orientational,tang2017orientation}. Determining the motility induced by these dipolar modes is complicated by the Stokes paradox and although this can be circumvented by various means we do not pursue this here. If such dipolar colloids were fixed within the material they would pump the ambient fluid and so it should be possible to use them to produce the concentration, filtering and corralling effects observed previously by funneling motile bacteria \cite{galajda2007wall}.

In line with our discussion at the beginning of this section, the basis quadrupoles are given by the real and imaginary parts of $\partial^2_z$, these being an achiral and chiral mode respectively, which are shown along with their flows in Fig.~\ref{fig:2DActiveMonoDi}. The flow generated by the achiral quadrupole in Fig.~\ref{fig:2DActiveMonoDi}(d) is purely radial and resembles the stresslet flow, unsurprising as it results from differentiating the vertical dipole in the same way as the stresslet is related to the Stokeslet. It is produced by a quadrupole distortion which may be associated with normal anchoring on the disc -- its counterpart with tangential anchoring has all the charges in its representative defect configuration inverted and a reversed flow response. Just as for the dipole distortions, the character of the quadrupole can be smoothly varied through adapting the boundary condition and the topological defects which represent the harmonic rotate rigidly in step with the changing anchoring angle. A generic anchoring angle will produce a net active torque, maximised for an angle of $\pi/4$ as illustrated for the chiral quadrupole shown in Fig.~\ref{fig:2DActiveMonoDi}(e). For extensile activity this distortion generates clockwise rotation, as can easily be justified via our representation of the far-field director structure as arising from a square arrangement of two $+1/2$ and two $-1/2$ defects -- the dual mode with the defect charges interchanged rotates anticlockwise. By choosing boundary conditions such that the defects are positioned closer to the mid-line of the colloid the strength of the active torque can be tuned.

\section{Chiral active stresses}
\label{sec:chiral_stresses}

Chirality is a ubiquitous trait, in living systems and liquid crystals alike. In active matter it opens a wealth of new phenomena, including odd viscous \cite{banerjee2017odd} and elastic responses \cite{scheibner2020odd,fruchart2022odd}, surface waves, rotating crystals \cite{tan2022odd} and non-reciprocal interactions \cite{bowick2022symmetry}. Chiral active stresses induce vortex arrays in active cholesterics~\cite{kole2021layered} and have also been shown to be important in nematic cell monolayers where they modify collective motion, the motility of topological defects and generate edge currents~\cite{hoffmann2020chiral,yashunsky2022chiral}. 
We now consider the effects of such chiral active stresses on nematic multipoles, both in two and three dimensions. 

\subsection{Two dimensions}

\begin{figure}
    \centering
    \includegraphics[width=0.55\textwidth]{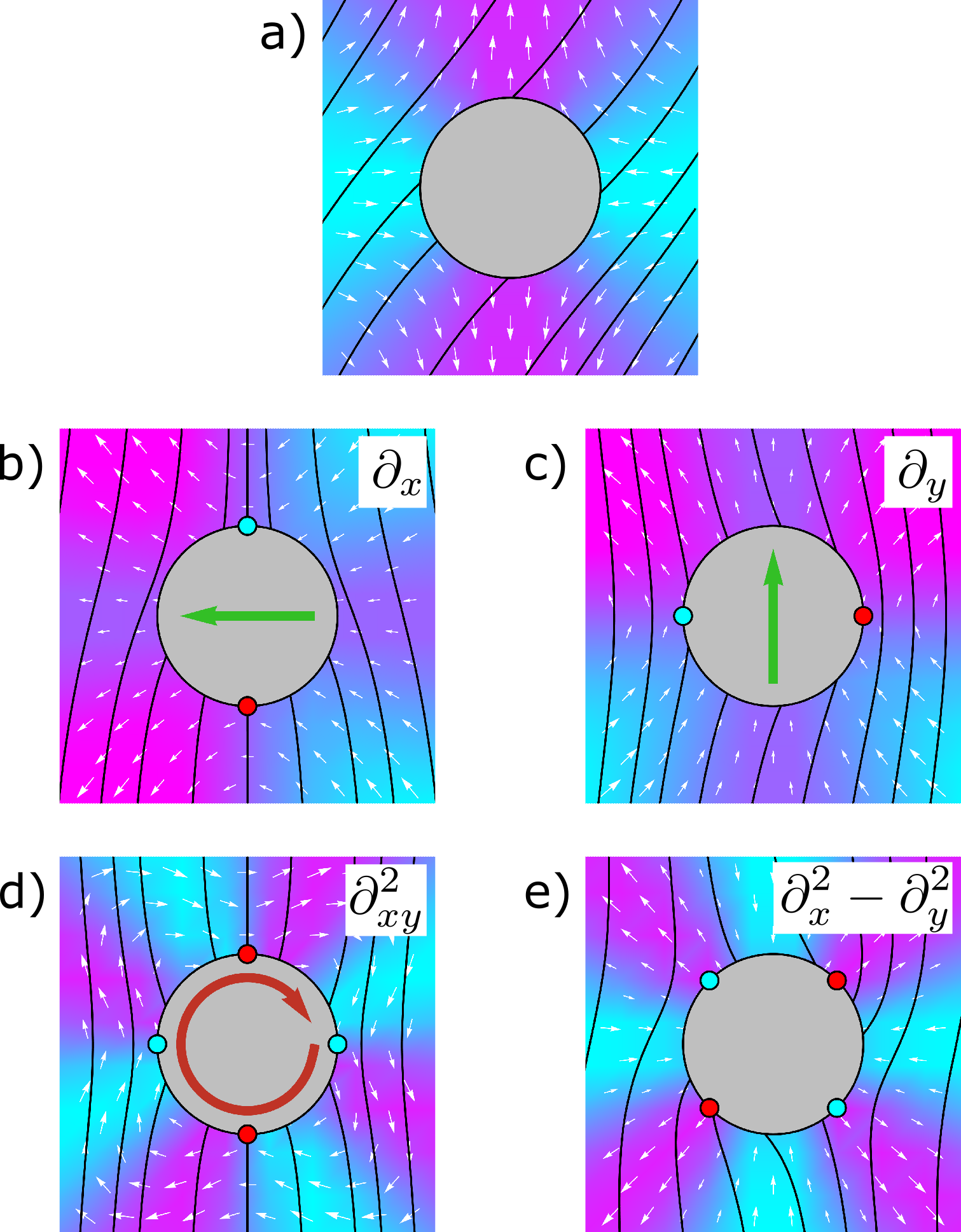}
    \caption[Distortions up to quadrupole order in two-dimensional chiral active nematics.]{Distortions up to quadrupole order in two-dimensional active nematics with purely chiral stresses. The active flow in white is superposed on the pressure field, with the integral curves of the director shown in black. (a) The fundamental monopole response is extensional and grows linearly with distance from the distortion. (b) and (c) show the flows induced by dipole distortions, labelled by the appropriate derivative of the nematic monopole, with the green arrows indicating the direction of self-propulsion that would result from net active forces in extensile systems.}
    \label{fig:2DActiveChiralStresses}
\end{figure}

For chiral stresses in two dimensions, the active stress tensor has the form $\boldsymbol{\sigma}^{\textrm{c}} = \chi J({\bf nn} - {\bf n}_{\perp}{\bf n}_{\perp})/2$, where $J$ is the complex structure defined by $J{\bf n} = {\bf n}_{\perp}$ and $J{\bf n}_{\perp} = -{\bf n}$. The chiral active force is 
\begin{equation}
    \nabla \cdot \boldsymbol{\sigma}^{\textrm{c}} = \chi J \Bigl( \nabla \cdot ({\bf nn}) \Bigr) ,
\end{equation}
and is simply a $\pi/2$ rotation of the achiral active force. Accordingly we can modify~\eqref{eq:StokesComplexAchiral} to give
\begin{equation}
    -p+i\mu\omega=-\frac{\zeta\theta_0}{2\log(a/R)}\frac{\bar{z}}{z} ,
\end{equation}
and solve as before to find
\begin{gather}
    \Tilde{{\bf u}} = \frac{\chi\theta_0}{8\mu\log(a/R)} \biggl[\frac{2xy}{r^2}(-y\mathbf{e}_x+x\mathbf{e}_y)+ 2\log\left(\frac{r}{R}\right)(-x\mathbf{e}_x+y\mathbf{e}_y)\biggr],  \\[2mm]
    \Tilde{p} = \frac{\chi \theta_0}{\log(a/R)} \frac{x^2-y^2}{2r^2}.
\end{gather}

Another way to understand the relation between achiral and chiral stresses is that, since the monopole active force field is spin-2, the $\pi/2$ local rotation of the active force results in a global rotation by $\pi/4$ of the force field and hence the fundamental flow responses. The action of this global rotation, denoted $R_{\pi/4}$, may be seen by comparing the monopole flow responses for achiral and chiral stresses, shown in Fig.~\ref{fig:2DActiveMonoDi}(a) and Fig.~\ref{fig:2DActiveChiralStresses}(a) respectively. For distortions of order $n$ there are two basis flows, $u_r$ and $u_i$, corresponding to the real and imaginary parts of $\partial_z^n$ respectively. The rotation of the monopole response has the consequence that for achiral and chiral active stresses these flows are related by
\begin{align}
    u_r^c&=R_{\pi/4}\left[\cos\left(\frac{n\pi}{4}\right)u_r^a-\sin\left(\frac{n\pi}{4}\right)u_i^a\right] ,\\
        u_i^c&=R_{\pi/4}\left[\sin\left(\frac{n\pi}{4}\right)u_r^a+\cos\left(\frac{n\pi}{4}\right)u_i^a\right],
\end{align}
where the superscripts denote the nature of the stresses as achiral or chiral. Hence flow solutions for chiral and achiral stresses are related by a clockwise rotation by $n\pi/4$ in the space of solutions followed by a rigid spatial rotation anticlockwise by $\pi/4$, as can be seen in Fig~\ref{fig:2DActiveChiralStresses}. At dipole order the chiral flow fields are rotated superpositions of the achiral ones, with the overall effect of chirality being to rotate the self-propulsion direction anticlockwise by $\pi/2$, interchanging the roles of horizontal and vertical propulsion. For a generic mixture of achiral and chiral stresses the direction of self-propulsion is rotated from the achiral case by an angle $\arctan(\chi/\zeta)$, mirroring the effect such stresses have on the flow profile of a $+1/2$ defect \cite{hoffmann2020chiral}. For the quadrupole distortions we have $u_i^c=R_{\pi/4}u_r^a$ and $u_i^c=R_{\pi/4}(-u_i^a)=u_i^a$, again swapping which distortion produces a chiral or achiral flow response. It is worth emphasising that the sign of the macroscopic rotation is not necessarily the same as the sign of the chiral stresses, rather it is the product of the signs of the activity and the distortion, just as for achiral stresses.

\subsection{Three dimensions}

In three dimensions the chiral active force is $\chi\nabla\times\left[\nabla\cdot(\mathbf{n}\mathbf{n})\right]$ \cite{kole2021layered} and so, by linearity, the fundamental flow responses are given by the curl of those derived earlier, namely
\begin{align}
    \mathbf{u}^{(x)}&=\frac{a\chi}{2\mu r^3}\left[-\mathbf{e}_x xy+\mathbf{e}_y(x^2-z^2)+\mathbf{e}_z yz\right],\\
    \mathbf{u}^{(y)}&=\frac{a\chi}{2\mu r^3}\left[-\mathbf{e}_x (y^2-z^2)+\mathbf{e}_y xy+-\mathbf{e}_z xz\right],
\end{align}
for monopole distortions in the $x$- and $y$-components respectively. Just as for achiral active stresses, we can combine these into a single complex fundamental flow response as $\mathbf{u}^{(x)}-i\mathbf{u}^{(y)}$, giving
\begin{equation}
    \Tilde{u}=\frac{i}{r^3}\left[-\bar{w}^2\mathbf{e}_w+(w\bar{w}-2z^2)\mathbf{e}_{\bar{w}}+2\bar{w}z\mathbf{e}_z\right].
\end{equation}
Since the active chiral force is a pure curl the corresponding pressure is constant.

\begin{figure}
    \centering
    \includegraphics[width=0.75\textwidth]{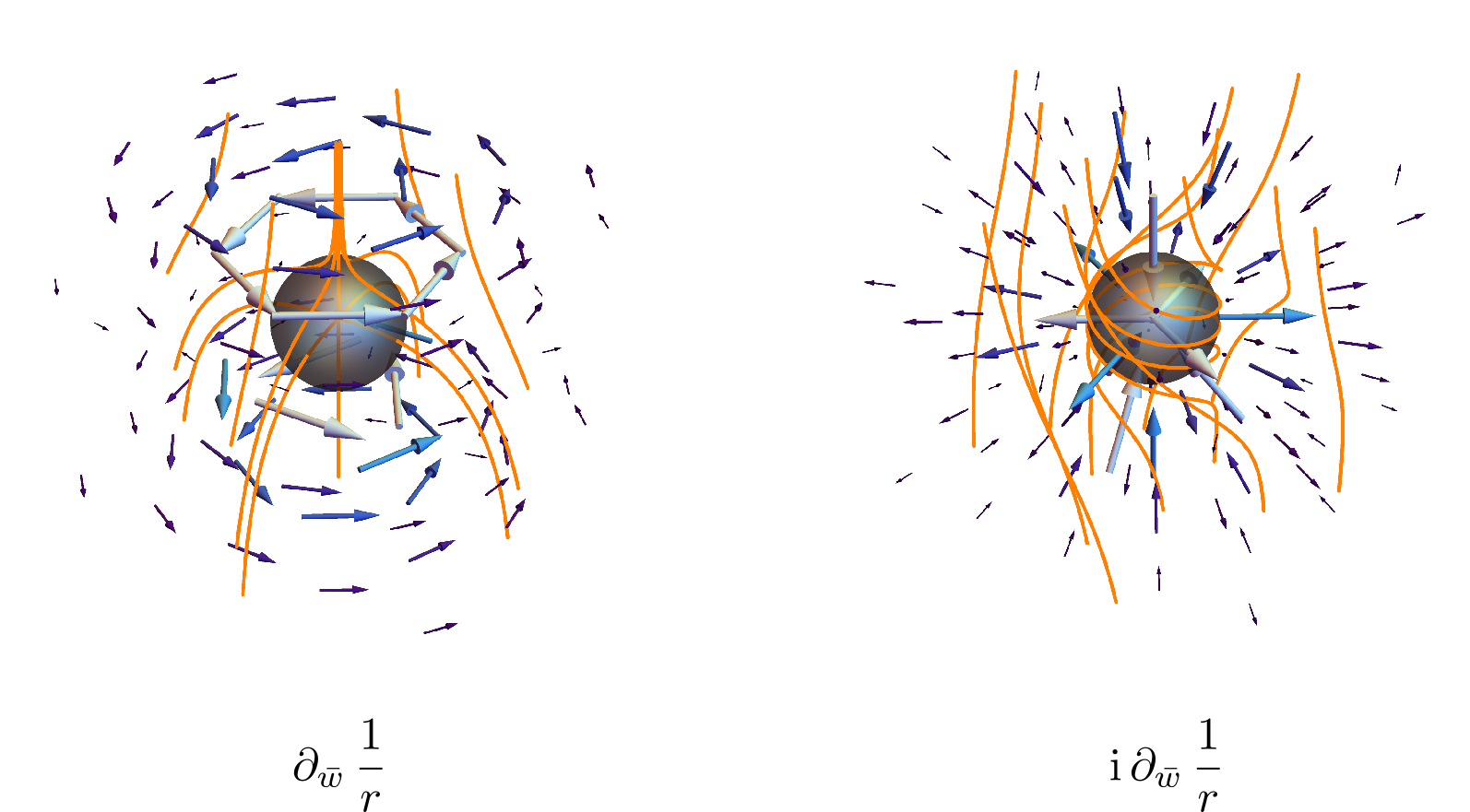}
    \caption[The active flows induced by spin $0$ dipole distortions with chiral active stresses.]{The active flows induced by spin $0$ dipole distortions with chiral active stresses. The flow is superposed upon the integral curves of the director, shown in orange, for the UPenn dipole (left) and chiral dipole (right).}
    \label{fig:ChiralStressesFlows3D}
\end{figure}

Owing to the additional derivative the functional behaviour of the flow responses is shifted up one order of distortion compared to achiral stresses, meaning dipole distortions induce rotations, although it should be noted that monopoles do not produce propulsive flows. The monopole flow responses are still spin-$1$, but since the flow response for a monopole distortion in $n_x$ for achiral stresses is primarily in the $x-z$ plane, the action of curl produces a flow that is dominantly in the $y$-direction and similarly the response to a monopole distortion in $n_y$ is mainly along $\mathbf{e}_x$. Together these ingredients mean that heuristically the flow response of a given distortion with chiral active stresses will resemble the achiral active stress flow response of the conjugate distortion at one higher order and with the same spin, that is the distortion reached by the action of $i\partial_z$. This is illustrated in Fig.~\ref{fig:ChiralStressesFlows3D} for the spin-$0$ dipoles. The UPenn dipole induces rotation about $\mathbf{e}_z$ while the chiral dipole produces a purely radial flow, resembling the achiral flow responses of the chiral quadrupole and Saturn's ring quadurpole respectively.

The phenomenological response can again be captured through integration of the stress tensor over a large sphere of radius $r$, just as was done for achiral active stresses. To enable us to reduce the active torque to a single boundary integral we use the symmetric form of the chiral active stress tensor~\cite{kole2021layered}, $\sigma_{ij}^c=\left[\nabla\times(\mathbf{n}\mathbf{n})\right]_{ij}+\left[\nabla\times(\mathbf{n}\mathbf{n})\right]_{ji}$, such that to linear order in director distortions we have
\begin{gather}
        \mathbf{f}^a = \int \chi {\bf \sigma^c} \cdot \text{d}{\bf A} \approx 0 , 
        \label{eq:active_force_chiral} \\[2mm]
        \begin{split}
            \boldsymbol{\tau}^a & = \int {\bf x} \times \chi {\bf \sigma^c} \cdot \text{d}{\bf A}  \approx \int \chi \biggl\{ {\bf e}_x \biggl[ \frac{xz \,\partial_x\delta n_x-2yz\partial_y\delta n_x+(y^2-z^2)\partial_z \delta n_x}{r} \biggr] \\
            & + {\bf e}_y \biggl[ \frac{yz \partial_y \delta n_y-2xz\partial_x \delta n_y +(x^2-z^2)\partial_z \delta n_y}{r} \biggr] \\
            & + {\bf e}_z \frac{-2xy(\partial_y \delta n_x+\partial_x \delta n_y)-(x^2-y^2)(\partial_x \delta n_x-\partial_y \delta n_y)+z(x\partial_z \delta n_x + y\partial_z \delta n_y) }{r} \biggr\} \text{d}A . 
        \end{split} \label{eq:active_torque_chiral}
    \end{gather}
From the first of these equations we see that, to linear order, there are no harmonic distortions which produce net forces in a nematic with chiral active stresses. With regard to the net active torques, the $x-$ and $y-$ components involve only $\delta n_x$ and $\delta n_y$ respectively and each term yields a non-zero integral only for $\delta n_i\sim\partial_z 1/r$, hence the two spin-$1$ dipoles produce transverse torques. Turning to the $z$-component, each term gives a non-zero integral only for $\delta n_i\sim\partial_i 1/r$, and as the expression is symmetric under interchange of $x$ and $y$ we see that only the UPenn dipole produces torques around $\mathbf{e}_z$. In other words, a dipolar director distortion which produces a net active force along a given direction in an achiral active nematic produces a net torque around the same direction in a chiral active nematic. These results of course accord with our earlier statements regarding the spins of distortions which are capable of producing torques about given axes. Performing the integrals we find that in each case the net active torque has magnitude $-12\pi\chi\alpha a^2/5$. Balancing this against Stokes drag gives, using the UPenn dipole as an example, an angular velocity
\begin{equation}
    \mathbf{\Omega}=\frac{3\chi\alpha}{10\mu a}\mathbf{e}_z.
\end{equation}
While the angular velocity in achiral active nematics is independent of the distortion size, in chiral active nematics it is inversely proportional to the radius, a direct consequence of the additional derivative in the active stress tensor. Accordingly, in chiral active nematics the rotational velocity is largest for smaller colloids.

\section{Discussion}

We have introduced active nematic multipoles as a novel framework for understanding the dynamics of active nematics. Although only formally valid on mesoscopic lengthscales, this approach produces results for the propulsive dynamics of defect loops that agree with those of a local analysis \cite{binysh2020three,houston2022defect}. It also provides various testable predictions, for example for the axis of self-propulsion or rotation induced by a distortion or how the corresponding velocities would scale with the size of a colloid.

More broadly, our results reveal self-propulsion and rotation as generic non-equilibrium responses that naturally arise due to colloidal inclusions in active nematics but also provide a template for the tailored design of particular dynamics. This provides insight into the issue of harnessing the energy of active systems to perform useful work, something which has been demonstrated in bacterial suspensions \cite{di2010bacterial,sokolov2010swimming} and is now receiving greater attention in the nematic context \cite{thampi2016active,zhang2021autonomous,loewe2022passive,yao2022topological}. Specific anchoring conditions on colloids have been investigated as a means of generating directed motion \cite{loewe2022passive}. Our results suggest that sufficient control of the anchoring conditions would allow for steerable and targeted colloidal delivery \cite{sahu2020omnidirectional}, although there may be routes to a similar degree of dynamical control through colloidal geometry alone \cite{MyThesis}. 

The transformative power of colloids in passive nematics was revealed in their collective behaviour, forming crystalline structures \cite{muvsevivc2006two,vskarabot2007two,vskarabot2008interactions,ognysta20082d,ognysta2011square} which can serve as photonic metamaterials \cite{muvsevivc2013nematic}. While our predictions for the dynamics of individual colloids have utility in their own right, there is again considerable interest in the collective dynamics which might emerge \cite{bricard2013emergence}. Although our results are insufficient to fully address these questions, some basic points can nonetheless be extracted from the flow solutions. The long-range nature of the active flows suggests that the hydrodynamic interactions will be dominant over elastic ones. The leading contribution to the pair-wise hydrodynamic interactions will be the advection of each colloid by the flow field generated by the other, and the even inversion symmetry of dipole flows implies that this provides a mechanism for pair-wise propulsion, even for colloids which are not self-propulsive themselves.

To conclude, it has been long-established that the distinct symmetries of $\pm 1/2$ nematic defects can be directly related to the qualitatively different dynamics they display in active systems \cite{giomi2013defect,giomi2014defect}. The aim of this paper is to bring the insights of this symmetry-based approach to generic nematic distortions.

\acknowledgements{This work was supported by the UK EPSRC through Grant No.~EP/N509796/1.}

%


\bibliographystyle{unsrt}
\bibliography{ActiveNematicMultipoles}

\begin{thebibliography}{10}

\bibitem{ramaswamy2010mechanics}
S.~Ramaswamy.
\newblock The mechanics and statistics of active matter.
\newblock {\em Annu. Rev. Condens. Matter Phys.}, 1(1):323--345, 2010.

\bibitem{marchetti2013hydrodynamics}
M.C. Marchetti, J.-F. Joanny, S.~Ramaswamy, T.B. Liverpool, J.~Prost, M.~Rao,
  and R.A. Simha.
\newblock Hydrodynamics of soft active matter.
\newblock {\em Rev. Mod. Phys.}, 85(3):1143, 2013.

\bibitem{doostmohammadi2018active}
A.~Doostmohammadi, J.~Ign{\'e}s-Mullol, J.M. Yeomans, and F.~Sagu{\'e}s.
\newblock Active nematics.
\newblock {\em Nat. Commun.}, 9(1):1--13, 2018.

\bibitem{duclos2017topological}
G.~Duclos, C.~Erlenk{\"a}mper, J.-F. Joanny, and P.~Silberzan.
\newblock Topological defects in confined populations of spindle-shaped cells.
\newblock {\em Nat. Phys.}, 13(1):58--62, 2017.

\bibitem{saw2017topological}
T.B. Saw, A.~Doostmohammadi, V.~Nier, L.~Kocgozlu, S.~Thampi, Y.~Toyama,
  P.~Marcq, C.T. Lim, J.M. Yeomans, and B.~Ladoux.
\newblock Topological defects in epithelia govern cell death and extrusion.
\newblock {\em Nature}, 544(7649):212--216, 2017.

\bibitem{zhou2014living}
S.~Zhou, A.~Sokolov, O.D. Lavrentovich, and I.S. Aranson.
\newblock Living liquid crystals.
\newblock {\em Proc. Natl. Acad. Sci. U.S.A.}, 111(4):1265--1270, 2014.

\bibitem{wensink2012meso}
H.H. Wensink, J.~Dunkel, S.~Heidenreich, K.~Drescher, R.E. Goldstein,
  H.~L{\"o}wen, and J.M. Yeomans.
\newblock Meso-scale turbulence in living fluids.
\newblock {\em Proc. Natl. Acad. Sci. U.S.A.}, 109(36):14308--14313, 2012.

\bibitem{sanchez2012spontaneous}
T.~Sanchez, D.T.N. Chen, S.J. DeCamp, M.~Heymann, and Z.~Dogic.
\newblock Spontaneous motion in hierarchically assembled active matter.
\newblock {\em Nature}, 491(7424):431--434, 2012.

\bibitem{adhyapak2013live}
T.C. Adhyapak, S.~Ramaswamy, and J.~Toner.
\newblock Live soap: stability, order, and fluctuations in apolar active
  smectics.
\newblock {\em Phys. Rev. Lett.}, 110(11):118102, 2013.

\bibitem{chen_toner2013}
L.~Chen and J.~Toner.
\newblock Universality for moving stripes: A hydrodynamic theory of polar
  active smectics.
\newblock {\em Phys. Rev. Lett.}, 111(088701), 2013.

\bibitem{whitfield2017hydrodynamic}
C.A. Whitfield, T.C. Adhyapak, A.~Tiribocchi, G.P. Alexander, D.~Marenduzzo,
  and S.~Ramaswamy.
\newblock Hydrodynamic instabilities in active cholesteric liquid crystals.
\newblock {\em Eur. Phys. J.E.}, 40(4):1--16, 2017.

\bibitem{kole2021layered}
S.J. Kole, G.P. Alexander, S.~Ramaswamy, and A.~Maitra.
\newblock Layered chiral active matter: Beyond odd elasticity.
\newblock {\em Phys. Rev. Lett.}, 126(24):248001, 2021.

\bibitem{maitra2020}
A.~Maitra, M.~Lenz, and R.~Voituriez.
\newblock Chiral active hexatics: Giant number fluctuations, waves, and
  destruction of order.
\newblock {\em Phys. Rev. Lett.}, 125(238005), 2020.

\bibitem{narayan2007}
V.~Narayan, S.~Ramaswamy, and N.~Menon.
\newblock Long-lived giant number fluctuations in a swarming granular nematic.
\newblock {\em Science}, 317:105, 2007.

\bibitem{giomi2013defect}
L.~Giomi, M.J. Bowick, X.~Ma, and M.C. Marchetti.
\newblock Defect annihilation and proliferation in active nematics.
\newblock {\em Phys. Rev. Lett.}, 110(22):228101, 2013.

\bibitem{giomi2014defect}
L.~Giomi, M.J. Bowick, P.~Mishra, R.~Sknepnek, and M.C. Marchetti.
\newblock Defect dynamics in active nematics.
\newblock {\em Phil. Trans. R. Soc. A}, 372(2029):20130365, 2014.

\bibitem{shendruk2017dancing}
T.N. Shendruk, A.~Doostmohammadi, K.~Thijssen, and J.M. Yeomans.
\newblock Dancing disclinations in confined active nematics.
\newblock {\em Soft Matter}, 13(21):3853--3862, 2017.

\bibitem{norton2018insensitivity}
M.M. Norton, A.~Baskaran, A.~Opathalage, B.~Langeslay, S.~Fraden, A.~Baskaran,
  and M.F. Hagan.
\newblock Insensitivity of active nematic liquid crystal dynamics to
  topological constraints.
\newblock {\em Phys. Rev. E}, 97(1):012702, 2018.

\bibitem{opathalage2019self}
A.~Opathalage, M.M. Norton, M.P.N. Juniper, B.~Langeslay, S.A. Aghvami,
  S.~Fraden, and Z.~Dogic.
\newblock Self-organized dynamics and the transition to turbulence of confined
  active nematics.
\newblock {\em Proc. Natl. Acad. Sci. U.S.A.}, 116(11):4788--4797, 2019.

\bibitem{maroudas-sacks2021}
Y.~Maroudas-Sacks, L.~Garion, L.~Shani-Zerbib, A.~Livshits, E.~Braun, and
  K.~Keren.
\newblock Topological defects in the nematic order of actin fibres as
  organization centres of {\it hydra} morphogenesis.
\newblock {\em Nat. Phys.}, 17:251, 2021.

\bibitem{duclos2020topological}
G.~Duclos, R.~Adkins, D.~Banerjee, M.S.E. Peterson, M.~Varghese, I.~Kolvin,
  A.~Baskaran, R.A. Pelcovits, T.R. Powers, A.~Baskaran, et~al.
\newblock Topological structure and dynamics of three-dimensional active
  nematics.
\newblock {\em Science}, 367(6482):1120--1124, 2020.

\bibitem{vcopar2019topology}
S.~{\v{C}}opar, J.~Aplinc, {\v{Z}}.~Kos, S.~{\v{Z}}umer, and M.~Ravnik.
\newblock Topology of three-dimensional active nematic turbulence confined to
  droplets.
\newblock {\em Phys. Rev. X}, 9(3):031051, 2019.

\bibitem{binysh2020three}
J.~Binysh, {\v{Z}}.~Kos, S.~{\v{C}}opar, M.~Ravnik, and G.P. Alexander.
\newblock Three-dimensional active defect loops.
\newblock {\em Phys. Rev. Lett.}, 124(8):088001, 2020.

\bibitem{houston2022defect}
A.J.H. Houston and G.P. Alexander.
\newblock Defect loops in three-dimensional active nematics as active
  multipoles.
\newblock {\em Phys. Rev. E}, 105(6):L062601, 2022.

\bibitem{poulin1997novel}
P.~Poulin, H.~Stark, T.C. Lubensky, and D.A. Weitz.
\newblock Novel colloidal interactions in anisotropic fluids.
\newblock {\em Science}, 275(5307):1770--1773, 1997.

\bibitem{stark2001physics}
H.~Stark.
\newblock Physics of colloidal dispersions in nematic liquid crystals.
\newblock {\em Phys. Rep.}, 351(6):387--474, 2001.

\bibitem{muvsevivc2017liquid}
I.~Mu{\v{s}}evi{\v{c}}.
\newblock {\em Liquid crystal colloids}.
\newblock Springer, 2017.

\bibitem{muvsevivc2006two}
I.~Mu{\v{s}}evi{\v{c}}, M.~{\v{S}}karabot, U.~Tkalec, M.~Ravnik, and
  S.~{\v{Z}}umer.
\newblock Two-dimensional nematic colloidal crystals self-assembled by
  topological defects.
\newblock {\em Science}, 313(5789):954--958, 2006.

\bibitem{ravnik2013confined}
M.~Ravnik and J.M. Yeomans.
\newblock Confined active nematic flow in cylindrical capillaries.
\newblock {\em Phys. Rev. Lett.}, 110(2):026001, 2013.

\bibitem{ravnik2007}
M.~Ravnik, M.~{\v{S}}karabot, S.~{\v{Z}}umer, U.~Tkalec, I.~Poberaj,
  D.~Babi{\v{c}}, N.~Osterman, and I.~Mu{\v{s}}evi{\v{c}}.
\newblock Entangled colloidal dimers and wires.
\newblock {\em Phys. Rev. Lett.}, 99(247801), 2007.

\bibitem{tkalec2011reconfigurable}
U.~Tkalec, M.~Ravnik, S.~{\v{C}}opar, S.~{\v{Z}}umer, and
  I.~Mu{\v{s}}evi{\v{c}}.
\newblock Reconfigurable knots and links in chiral nematic colloids.
\newblock {\em Science}, 333(6038):62--65, 2011.

\bibitem{guillamat2018}
P.~Guillamat, {\v{Z}}.~Kos, J.~Hardo\"uin, J.~Ign\'es-Mullol, M.~Ravnik, and
  F.~Sagu\'es.
\newblock Active nematic emulsions.
\newblock {\em Sci. Adv.}, 4(eaa01470), 2018.

\bibitem{hardouin2019}
J.~Hardo\"uin, P.~Guillamat, F.~Sagu\'es, and J.~Ign\'es-Mullol.
\newblock Dynamics of ring disclinations driven by active nematic shells.
\newblock {\em Front. Phys.}, 7(00165), 2019.

\bibitem{baek2018generic}
Y.~Baek, A.P. Solon, X.~Xu, N.~Nikola, and Y.~Kafri.
\newblock Generic long-range interactions between passive bodies in an active
  fluid.
\newblock {\em Phys. Rev. Lett.}, 120(5):058002, 2018.

\bibitem{rajabi2020directional}
M.~Rajabi, H.~Baza, T.~Turiv, and O.D. Lavrentovich.
\newblock Directional self-locomotion of active droplets enabled by nematic
  environment.
\newblock {\em Nat. Phys.}, pages 1--7, 2020.

\bibitem{loewe2022passive}
B.~Loewe and T.N. Shendruk.
\newblock Passive janus particles are self-propelled in active nematics.
\newblock {\em New J. Phys.}, 24(1):012001, 2022.

\bibitem{yao2022topological}
T.~Yao, {\v{Z}}.~Kos, Q.X. Zhang, Y.~Luo, E.B. Steager, M.~Ravnik, and K.J.
  Stebe.
\newblock Topological defect-propelled swimming of nematic colloids.
\newblock {\em Sci. Adv.}, 8(34):eabn8176, 2022.

\bibitem{frank1958liquid}
F.C. Frank.
\newblock I. liquid crystals. on the theory of liquid crystals.
\newblock {\em Discuss. Faraday Soc.}, 25:19--28, 1958.

\bibitem{angheluta2021role}
L.~Angheluta, Z.~Chen, M.C. Marchetti, and M.J. Bowick.
\newblock The role of fluid flow in the dynamics of active nematic defects.
\newblock {\em New J. Phys.}, 23(3):033009, 2021.

\bibitem{khoromskaia2017vortex}
D.~Khoromskaia and G.P. Alexander.
\newblock Vortex formation and dynamics of defects in active nematic shells.
\newblock {\em New J. Phys.}, 19(10):103043, 2017.

\bibitem{alert2020universal}
R.~Alert, J.-F. Joanny, and J.~Casademunt.
\newblock Universal scaling of active nematic turbulence.
\newblock {\em Nat. Phys.}, 16(6):682--688, 2020.

\bibitem{simha2002hydrodynamic}
R.A. Simha and S.~Ramaswamy.
\newblock Hydrodynamic fluctuations and instabilities in ordered suspensions of
  self-propelled particles.
\newblock {\em Phys. Rev. Lett.}, 89(5):058101, 2002.

\bibitem{maxwell1873treatise}
J.C. Maxwell.
\newblock {\em A treatise on electricity and magnetism}, volume~1.
\newblock Oxford: Clarendon Press, 1873.

\bibitem{dennis2004canonical}
M.R. Dennis.
\newblock Canonical representation of spherical functions: Sylvester's theorem,
  maxwell's multipoles and majorana's sphere.
\newblock {\em J. Phys. A Math.}, 37(40):9487, 2004.

\bibitem{arnold1996topological}
V.~Arnold et~al.
\newblock Topological content of the maxwell theorem on multiple representation
  of spherical functions.
\newblock {\em Topol. Methods Nonlinear Anal.}, 7(2):205--217, 1996.

\bibitem{brochard1970theory}
F.~Brochard and P.-G. De~Gennes.
\newblock Theory of magnetic suspensions in liquid crystals.
\newblock {\em Journal de Physique}, 31(7):691--708, 1970.

\bibitem{yuan2019elastic}
Y.~Yuan, Q.~Liu, B.~Senyuk, and I.I. Smalyukh.
\newblock Elastic colloidal monopoles and reconfigurable self-assembly in
  liquid crystals.
\newblock {\em Nature}, 570(7760):214--218, 2019.

\bibitem{lubensky1998topological}
T.C. Lubensky, D.~Pettey, N.~Currier, and H.~Stark.
\newblock Topological defects and interactions in nematic emulsions.
\newblock {\em Phys. Rev. E}, 57(1):610, 1998.

\bibitem{pergamenshchik2011dipolar}
V.M. Pergamenshchik and V.A. Uzunova.
\newblock Dipolar colloids in nematostatics: tensorial structure, symmetry,
  different types, and their interaction.
\newblock {\em Phys. Rev. E}, 83(2):021701, 2011.

\bibitem{terentjev1995disclination}
E.M. Terentjev.
\newblock Disclination loops, standing alone and around solid particles, in
  nematic liquid crystals.
\newblock {\em Phys. Rev. E}, 51(2):1330, 1995.

\bibitem{tkalec2009}
U.~Tkalec, M.~Ravnik, S.~{\v{Z}}umer, and I.~, Mu{\v{s}}evi{\v{c}}.
\newblock Vortexlike topological defects in nematic colloids: Chiral colloidal
  dimers and 2d crystals.
\newblock {\em Phys. Rev. Lett.}, 103(127801), 2009.

\bibitem{alexander2018topology}
G.P. Alexander.
\newblock Topology in liquid crystal phases.
\newblock In {\em The Role of Topology in Materials}, pages 229--257. Springer,
  2018.

\bibitem{smalyukh2010three}
I.I. Smalyukh, Y.~Lansac, N.A. Clark, and R.P. Trivedi.
\newblock Three-dimensional structure and multistable optical switching of
  triple-twisted particle-like excitations in anisotropic fluids.
\newblock {\em Nat. Mater.}, 9(2):139--145, 2010.

\bibitem{ravnik2009landau}
M.~Ravnik and S.~{\v{Z}}umer.
\newblock Landau--de gennes modelling of nematic liquid crystal colloids.
\newblock {\em Liquid Crystals}, 36(10-11):1201--1214, 2009.

\bibitem{blake1974fundamental}
J.R. Blake and A.T. Chwang.
\newblock Fundamental singularities of viscous flow.
\newblock {\em J. Eng. Math.}, 8(1):23--29, 1974.

\bibitem{chwang1975hydromechanics}
A.T. Chwang and T.Y.T. Wu.
\newblock Hydromechanics of low-reynolds-number flow. part 2. singularity
  method for stokes flows.
\newblock {\em J. Fluid Mech.}, 67(4):787--815, 1975.

\bibitem{khoromskaia2015motility}
D.~Khoromskaia and G.P. Alexander.
\newblock Motility of active fluid drops on surfaces.
\newblock {\em Phys. Rev. E}, 92(6):062311, 2015.

\bibitem{lighthill1952squirming}
M.J. Lighthill.
\newblock On the squirming motion of nearly spherical deformable bodies through
  liquids at very small reynolds numbers.
\newblock {\em Commun. Pure Appl. Math.}, 5(2):109--118, 1952.

\bibitem{pak2014generalized}
O.S. Pak and E.~Lauga.
\newblock Generalized squirming motion of a sphere.
\newblock {\em J. Eng. Math.}, 88(1):1--28, 2014.

\bibitem{daddi2018dynamics}
A.~Daddi-Moussa-Ider and A.M. Menzel.
\newblock Dynamics of a simple model microswimmer in an anisotropic fluid:
  Implications for alignment behavior and active transport in a nematic liquid
  crystal.
\newblock {\em Phys. Rev. Fluids}, 3(9):094102, 2018.

\bibitem{kos2018elementary}
{\v{Z}}.~Kos and M.~Ravnik.
\newblock Elementary flow field profiles of micro-swimmers in weakly
  anisotropic nematic fluids: Stokeslet, stresslet, rotlet and source flows.
\newblock {\em Fluids}, 3(1):15, 2018.

\bibitem{MyThesis}
A.J.H. Houston.
\newblock {\em Active and passive nematic multipoles}.
\newblock PhD thesis, University of Warwick, 2022.

\bibitem{binysh2018maxwell}
J.~Binysh and G.P. Alexander.
\newblock Maxwell’s theory of solid angle and the construction of knotted
  fields.
\newblock {\em J. Phys. A Math.}, 51(38):385202, 2018.

\bibitem{friedel1969boucles}
J.~Friedel and P.-G. De~Gennes.
\newblock Boucles de disclination dans les cristaux liquides.
\newblock {\em CR Acad. Sc. Paris B}, 268:257--259, 1969.

\bibitem{kim1991microhydrodynamics}
S.~Kim and S.J. Karrila.
\newblock Microhydrodynamics: Principles and selected applications, 1991.

\bibitem{loudet2004stokes}
J.C. Loudet, P.~Hanusse, and P.~Poulin.
\newblock Stokes drag on a sphere in a nematic liquid crystal.
\newblock {\em Science}, 306(5701):1525--1525, 2004.

\bibitem{ruhwandl1996friction}
R.W. Ruhwandl and E.M. Terentjev.
\newblock Friction drag on a particle moving in a nematic liquid crystal.
\newblock {\em Phys. Rev. E}, 54(5):5204, 1996.

\bibitem{stark2001stokes}
H.~Stark and D.~Ventzki.
\newblock Stokes drag of spherical particles in a nematic environment at low
  ericksen numbers.
\newblock {\em Phys. Rev. E}, 64(3):031711, 2001.

\bibitem{batchelor1971stress}
G.K. Batchelor.
\newblock The stress generated in a non-dilute suspension of elongated
  particles by pure straining motion.
\newblock {\em J. Fluid Mech.}, 46(4):813--829, 1971.

\bibitem{lang2019effects}
C.~Lang, J.~Hendricks, Z.~Zhang, N.K. Reddy, J.P. Rothstein, M.P. Lettinga,
  J.~Vermant, and C.~Clasen.
\newblock Effects of particle stiffness on the extensional rheology of model
  rod-like nanoparticle suspensions.
\newblock {\em Soft matter}, 15(5):833--841, 2019.

\bibitem{sokolov2010swimming}
A.~Sokolov, M.M. Apodaca, B.A. Grzybowski, and I.S. Aranson.
\newblock Swimming bacteria power microscopic gears.
\newblock {\em Proc. Natl. Acad. Sci. U.S.A.}, 107(3):969--974, 2010.

\bibitem{vromans2016orientational}
A.J. Vromans and L.~Giomi.
\newblock Orientational properties of nematic disclinations.
\newblock {\em Soft matter}, 12(30):6490--6495, 2016.

\bibitem{tang2017orientation}
X.~Tang and J.V. Selinger.
\newblock Orientation of topological defects in 2d nematic liquid crystals.
\newblock {\em Soft matter}, 13(32):5481--5490, 2017.

\bibitem{galajda2007wall}
P.~Galajda, J.~Keymer, P.~Chaikin, and R.~Austin.
\newblock A wall of funnels concentrates swimming bacteria.
\newblock {\em J. Bacteriol.}, 189(23):8704--8707, 2007.

\bibitem{banerjee2017odd}
D.~Banerjee, A.~Souslov, A.G. Abanov, and V.~Vitelli.
\newblock Odd viscosity in chiral active fluids.
\newblock {\em Nat. Commun.}, 8(1):1--12, 2017.

\bibitem{scheibner2020odd}
C.~Scheibner, A.~Souslov, D.~Banerjee, P.~Sur{\'o}wka, W.~Irvine, and
  V.~Vitelli.
\newblock Odd elasticity.
\newblock {\em Nat. Phys.}, 16(4):475--480, 2020.

\bibitem{fruchart2022odd}
M.~Fruchart, C.~Scheibner, and V.~Vitelli.
\newblock Odd viscosity and odd elasticity.
\newblock {\em arXiv preprint arXiv:2207.00071}, 2022.

\bibitem{tan2022odd}
T.H. Tan, A.~Mietke, J.~Li, Y.~Chen, H.~Higinbotham, P.J. Foster, S.~Gokhale,
  J.~Dunkel, and N.~Fakhri.
\newblock Odd dynamics of living chiral crystals.
\newblock {\em Nature}, 607(7918):287--293, 2022.

\bibitem{bowick2022symmetry}
M.J. Bowick, N.~Fakhri, M.C. Marchetti, and S.~Ramaswamy.
\newblock Symmetry, thermodynamics, and topology in active matter.
\newblock {\em Phys. Rev. X}, 12(1):010501, 2022.

\bibitem{hoffmann2020chiral}
L.A. Hoffmann, K.~Schakenraad, R.M.H. Merks, and L.~Giomi.
\newblock Chiral stresses in nematic cell monolayers.
\newblock {\em Soft matter}, 16(3):764--774, 2020.

\bibitem{yashunsky2022chiral}
V.~Yashunsky, D.J.G. Pearce, C.~Blanch-Mercader, F.~Ascione, P.~Silberzan, and
  L.~Giomi.
\newblock Chiral edge current in nematic cell monolayers.
\newblock {\em Phys. Rev. X}, 12(4):041017, 2022.

\bibitem{di2010bacterial}
R.~Di~Leonardo, L.~Angelani, D.~Dell’Arciprete, G.~Ruocco, V.~Iebba,
  S.~Schippa, M.P. Conte, F.~Mecarini, F.~De~Angelis, and E.~Di~Fabrizio.
\newblock Bacterial ratchet motors.
\newblock {\em Proc. Natl. Acad. Sci. U.S.A.}, 107(21):9541--9545, 2010.

\bibitem{thampi2016active}
S.P. Thampi, A.~Doostmohammadi, T.N. Shendruk, R.~Golestanian, and J.M.
  Yeomans.
\newblock Active micromachines: Microfluidics powered by mesoscale turbulence.
\newblock {\em Sci. Adv.}, 2(7):e1501854, 2016.

\bibitem{zhang2021autonomous}
R.~Zhang, A.~Mozaffari, and J.J. de~Pablo.
\newblock Autonomous materials systems from active liquid crystals.
\newblock {\em Nat. Rev. Mater.}, pages 1--17, 2021.

\bibitem{sahu2020omnidirectional}
D.K. Sahu, S.~Kole, S.~Ramaswamy, and S.~Dhara.
\newblock Omnidirectional transport and navigation of janus particles through a
  nematic liquid crystal film.
\newblock {\em Phys. Rev. Res.}, 2(3):032009, 2020.

\bibitem{vskarabot2007two}
M.~{\v{S}}karabot, M.~Ravnik, S.~{\v{Z}}umer, U.~Tkalec, I.~Poberaj,
  D.~Babi{\v{c}}, N.~Osterman, and I.~Mu{\v{s}}evi{\v{c}}.
\newblock Two-dimensional dipolar nematic colloidal crystals.
\newblock {\em Phys. Rev. E}, 76(5):051406, 2007.

\bibitem{vskarabot2008interactions}
M.~{\v{S}}karabot, M.~Ravnik, S.~{\v{Z}}umer, U.~Tkalec, I.~Poberaj,
  D.~Babi{\v{c}}, N.~Osterman, and I.~Mu{\v{s}}evi{\v{c}}.
\newblock Interactions of quadrupolar nematic colloids.
\newblock {\em Phys. Rev. E}, 77(3):031705, 2008.

\bibitem{ognysta20082d}
U.~Ognysta, A.~Nych, V.~Nazarenko, I.~Mu{\v{s}}evi{\v{c}}, M.~{\v{S}}karabot,
  M.~Ravnik, S.~{\v{Z}}umer, I.~Poberaj, and D.~Babi{\v{c}}.
\newblock 2d interactions and binary crystals of dipolar and quadrupolar
  nematic colloids.
\newblock {\em Phys. Rev. Lett.}, 100(21):217803, 2008.

\bibitem{ognysta2011square}
U.M. Ognysta, A.B. Nych, V.A. Uzunova, V.M. Pergamenschik, V.G. Nazarenko,
  M.~{\v{S}}karabot, and I.~Mu{\v{s}}evi{\v{c}}.
\newblock Square colloidal lattices and pair interaction in a binary system of
  quadrupolar nematic colloids.
\newblock {\em Phys. Rev. E}, 83(4):041709, 2011.

\bibitem{muvsevivc2013nematic}
I.~Mu{\v{s}}evi{\v{c}}.
\newblock Nematic colloids, topology and photonics.
\newblock {\em Phil. Trans. R. Soc. A}, 371(1988):20120266, 2013.

\bibitem{bricard2013emergence}
A.~Bricard, J.-B. Caussin, N.~Desreumaux, O.~Dauchot, and D.~Bartolo.
\newblock Emergence of macroscopic directed motion in populations of motile
  colloids.
\newblock {\em Nature}, 503(7474):95--98, 2013.

\end{thebibliography}

\end{document}